\documentclass[12pt, leqno, a4paper]{article}
\usepackage[toc,page]{appendix}
\usepackage{amsmath,setspace,multirow,lineno}
\usepackage[font=small,labelfont=bf]{caption}
\usepackage[top=1.3in, bottom=1.3in, left=1in, right=1in]{geometry}
\usepackage[round]{natbib}
\usepackage{color, soul, bbm}

\DeclareCaptionStyle{italic}[justification=centering]{labelfont={bf},textfont={it},labelsep=colon}
\usepackage{graphicx,psfrag,epsf,textcomp,epstopdf, amsthm}
\usepackage{enumerate, dsfont}
\usepackage{natbib,url}
\usepackage[table]{xcolor}
\usepackage{booktabs, subfig, bm, paralist, mathpazo, tikz, longtable, microtype}
\usepackage{algorithm}
\usepackage{float}
\usepackage{algpseudocode}

\usepackage[pdftex, colorlinks=true]{hyperref}
\definecolor{darkblue}{rgb}{0,0,.6}
\definecolor{skyblue}{RGB}{135,206,235}
\hypersetup{citecolor=darkblue,linkcolor=darkblue,urlcolor=darkblue}
\definecolor{DarkRed}{rgb}{.7,0,.4}
\definecolor{ao(english)}{rgb}{0.0, 0.5, 0.0}

\usepackage{comment,orcidlink}
\usepackage{xurl}
\DeclareMathAlphabet\mathbfcal{OMS}{cmsy}{b}{n}

\newcommand{\blind}{0}

\newcommand{\X}{\mathcal{X}}
\newcommand{\Y}{\mathcal{Y}}

\graphicspath{{plots/}}
\DeclareMathOperator*{\argmin}{\arg\!\min}

\newcommand{\Rlogo}{\protect\includegraphics[height=1.8ex,keepaspectratio]{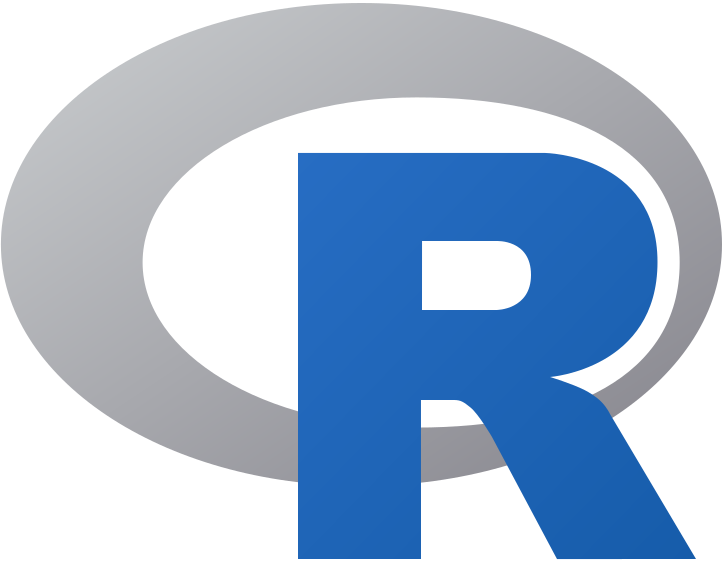}}

\newsavebox\CBox
\def\textBF#1{\sbox\CBox{#1}\resizebox{\wd\CBox}{\ht\CBox}{\textbf{#1}}}

\renewcommand\X{\mathcal{X}}

\newtheorem{@definition}{\sc Definition}[section]

\usepackage{array}
\newcommand{\PreserveBackslash}[1]{\let\temp=\\#1\let\\=\temp}
\newcolumntype{C}[1]{>{\PreserveBackslash\centering}p{#1}}
\newcolumntype{R}[1]{>{\PreserveBackslash\raggedleft}p{#1}}
\newcolumntype{L}[1]{>{\PreserveBackslash\raggedright}p{#1}}

\begin{document}

\def\spacingset#1{\renewcommand{\baselinestretch}{#1}\small\normalsize} \spacingset{1}

\if0\blind
{
\title{\bf Bias correction of long-memory estimator of functional time series via the prefiltered sieve bootstrap}}
\author{
\normalsize Chang Liu \orcidlink{0009-0007-6909-3195} \qquad Han Lin Shang \orcidlink{0000-0003-1769-6430}\footnote{Corresponding address: Department of Actuarial Studies and Business Analytics, Macquarie University, Sydney, NSW 2109, Australia; Telephone: +61(2) 9850 4689; Email: hanlin.shang@mq.edu.au} \\
\normalsize Department of Actuarial Studies and Business Analytics \\
\normalsize Macquarie University
}
\date{}
\maketitle
\fi

\if1\blind
{
\title{\bf Bias correction of long-memory estimator of functional time series via the prefiltered sieve bootstrap}
} \fi

\date{}
\maketitle

\begin{abstract}
We investigate a bias correction procedure based on sieve bootstrapping to estimate the long-memory parameter $d$ in stationary or nonstationary fractionally integrated processes. The resampling method implements a sieve bootstrap method on data prefiltered by a preliminary estimate of the long-memory parameter. For the initial estimate, we recommend the local polynomial Whittle with noise (LPWN) estimator in \cite{FNN12} to reduce bias, especially in the presence of a strong short-range autoregressive dependence. Through a series of simulation studies, we first highlight the issue of bias, using the local Whittle or detrended fluctuation analysis estimator. Then, we consider the LPWN estimator and show the potential improvement in bias achieved by its sieve bootstrap enhancement. As a byproduct, the sieve bootstrap can also provide confidence intervals of the memory parameter.

\vspace{.05in}
\end{abstract}

\newpage
\setstretch{1.5}

\section{Introduction}

Recent advances in computing and storage have facilitated the presence of functional data with graphical representations of curves \citep{RS05}, images \citep{SGS13}, shapes \citep{SK16}, and manifolds \citep{RC25}. Differing from multivariate data, each datum in functional data analysis is a random function that resides in a metric space, commonly a separable Hilbert space. For a comprehensive overview of first- and second-generation functional data, we refer to \cite{RS05} and \cite{KS23}, respectively. Second-generation functional data include the analysis of functional time series, which balances functional data and time series analyses.

Similarly to its univariate and multivariate counterparts, a functional time series exhibits temporal dependence in its observations. There is a growing body of literature on the development of statistical methods and theory to model temporal dependence within a stationary and (less common) nonstationary framework. The bulk of the literature assumes short-range dependence. These include some parametric tools, such as the functional autoregressive (FAR) model \citep{Bosq00, KR13}, the functional moving average (FMA) model \citep{KK17}, and the functional autoregressive moving average (FARMA) model \citep{KKW17}; and nonparametric tools, such as nonparametric functional regression \citep{FV06} and the functional factor model \citep{BYZ10, LLS+26}.

Moving beyond short-range dependence, we enter the realm of long-range dependence, which has wide practical relevance across scientific fields, from maturity-specific yield curves to age-specific demography. Also coined as long memory by \cite{Granger80} and \cite{GJ80}, such series describe processes with greater persistence than short-range dependent ones. In the stationary long-memory setting, autocovariance decays very slowly, and the spectral density is unbounded, typically at zero frequency. Since a fractionally integrated model can produce long memory, \cite{LRS20} proposed the functional autoregressive fractionally integrated moving average (FARFIMA) model, and \cite{SK23} introduced the tempered functional autoregressive fractionally integrated moving average (TFARFIMA) model. In the nonstationary setting, \cite{LRS23} proposed an iterative estimation algorithm to estimate the long-memory parameter via integer-order differencing.

Long-memory functional time series also give rise to the concept of cointegration \citep[see, e.g.,][]{CKP16}. \cite{BSS17} and \cite{SB19} extended the Granger-Johansen representation theorem for nonstationary functional time series taking values in a Hilbert space and a Bayes Hilbert space, respectively. \cite{SS26} first developed a functional KPSS-type test to determine whether the time series of interest is $I(0)$ (i.e., short-range dependence) or $I(d)$ for $d\in (-0.5, 0.5)\backslash\{0\}$ (i.e., fractionally integrated). They then extended this KPSS-type test to determine whether a curve time series is $I(1)$ or integrated of any integer order; hence, appropriately differenced sequences are assumed to be stationary $I(0)$ sequences.

Estimating the strength of long-memory dependence is critical in analysing and modelling long-memory curve time series. The long-memory parameter controls the shape of the spectrum near the zero frequency and the hyperbolic decay of the series's autocorrelation function. For determining the memory parameter, two measures are commonly used: $H$, known as the Hurst exponent or the self-similarity parameter \citep{MN68}, and the fractional integration parameter~$d$, arising from a generalisation of autoregressive fractionally integrated moving average models from integer to noninteger values of the integration parameter $d$.

In the functional time-series literature, several papers have considered the estimation of the long-memory parameter $d$, see, e.g., the range/scale estimator of \cite{LRS20}, Detrended Fluctuation Analysis (DFA) of \cite{HNZ26}, and the local Whittle (LW) estimator of \cite{LRS21, LRS23}. A finite-sample comparison of bias and mean squared error among several estimators is presented in \cite{Shang22}, which recommends the DFA and LW among 12 estimators to estimate the memory parameter under the FARFIMA(1, $d$, 0) and FARFIMA(1, $d$, 1) data-generating processes, respectively. In the presence of strong dependencies on autoregressive coefficients, these estimators are prone to finite-sample bias, which can be quite severe. Our contribution is to provide a bias-correction procedure based on an initial bias-corrected estimator and sieve bootstrapping. 

The outline of this paper is presented as follows: In Section~\ref{sec:2}, we review the local Whittle estimator of \cite{RM95} and present a bias-corrected long-memory estimator of \cite{FNN12}. Following the early work of \cite{PMG17}, we introduce the bias correction via a sieve bootstrap idea of \cite{Paparoditis18}. In Section~\ref{sec:3} and \ref{sec:4}, we describe the sieve bootstrap procedure of \cite{Paparoditis18} and show how it can be used to improve bias correction. Through a series of simulation studies in Section~\ref{sec:5}, we consider FARFIMA(1, $d$, 0), FARFIMA(0, $d$, 1), and FARFIMA(1, $d$, 1) processes, with which we evaluate and compare the finite-sample estimation accuracy among the LW, local polynomial Whittle with noise (LPWN), and sieve-bootstrapped LPWN estimators. In Section~\ref{sec:6}, we apply these estimators to two empirical datasets, namely the Swedish age-specific mortality rates and Canadian yield curves. The conclusion is given in Section~\ref{sec:6}, along with some ideas on how the methodology presented can be further extended.

\section{Estimation of long-memory parameter}\label{sec:2}

Let $\Y_t(u)$ be a long-range dependent stationary or nonstationary time series of functions observed on a function support $u\in \mathcal{I}\subset R$, in which $\mathcal{I}$ is a subset of real space. From a finite sample $\bm{\Y}(u) = \{\Y_1(u),\dots,\Y_{T}(u)\}$ and $T$ denotes a sample size, we compute its sample covariance $\widehat{C}(u,v)$, from which we estimate its leading principal component and its associated principal component scores, defined as $\{\beta_t\}_{t=1}^{T}$. The memory parameter~$d$ is then estimated from this univariate time series using time-domain or frequency-domain methods \citep[see, e.g.,][for a comprehensive review]{Shang22}. In Sections~\ref{sec:2.1} and~\ref{sec:2.2}, we revisit the DFA estimator and the LW estimator, respectively. In the presence of strong short-range dependence, the LPWN estimator in Section~\ref{sec:2.3} was developed on the basis of the LW estimator to reduce bias. 

\subsection{Detrended Fluctuation Analysis Estimator}\label{sec:2.1}

Introduced by \cite{PBH+94} in biology, the DFA estimator is a regression-based estimator for studying the long-memory parameter in deoxyribonucleic acid sequences. The idea is to remove local trends at different scales and examine how variance changes with different block sizes. The scaling relationship between the variance and the blocks provides an estimate of the Hurst exponent (often denoted by $H$), from which the memory parameter $(d=H-\frac{1}{2})$ can be derived.
The estimator can be summarised with the following steps:
\begin{enumerate}
\item[1)]  Divide the univariate time series $(\beta_{1}, \beta_{2},\dots,\beta_{T})$ into $K$ nonoverlapping blocks with block size~$m$ such that $T=mK$.
\item[2)] For the $k$\textsuperscript{th} block, regress $T_{l,k} = \sum^l_{t=(k-1)m+1}\beta_t$ against $l$, where $l=mk$, $k=1,2,\dots,K$, and estimate the variance of the residuals by
\begin{equation*}
S_m(k) = \frac{1}{m}\sum^{km}_{l=(k-1)m+1}(T_{l,k}-\vartheta_{0,k}-\vartheta_{1,k}l)^2,
\end{equation*}
where $\vartheta_{0,k}$ and $\vartheta_{1,k}$ are ordinary least-squares estimates based on the $k$\textsuperscript{th} block.
\item[3)] Compute the average of the variance of the residuals
\begin{equation}
    \overline{S}_{m}^2 = \frac{1}{K}\sum^K_{k=1}S_m^2(k).\label{eq:2.1}
\end{equation}
\item[4)] For a range of block size values $m=1,2,\dots,M$, we evaluate $\overline{S}_{m}^2$ from~\eqref{eq:2.1}, and then regress $\log_{10}(\overline{S}_m^2)$ on $\log_{10}(m)$ to obtain the estimated regression coefficient $\widehat{\theta}$. The estimated value of the Hurst exponent is given by $\widehat{H} = \widehat{\theta}/2$, and the memory parameter $\widehat{d} = \widehat{H}-\frac{1}{2}$. 
\end{enumerate}

Although our study focuses on the monofractal memory parameter, the DFA estimator can be used to estimate multifractal memory parameters at various scales \citep[see, e.g.,][]{PYC21, HNZ26}.

\subsection{Local Whittle Estimator} \label{sec:2.2}

In addition to the DFA estimator, we briefly revisit some related frequency-domain estimators for~$d$, particularly the standard LW estimator and the LPWN estimator in Section~\ref{sec:2.3}. These estimators share the same Whittle-type objective function but differ in the approximation used for the spectral density near frequency zero.

Let $\lambda_j = (2\pi j)/T$ denote the Fourier frequencies, where $j=1, 2,\ldots,J$, and $J$ is a bandwidth parameter in the frequency domain satisfying $J \to \infty$ and $J \sim o(T)$ as $T \to \infty$. The discrete Fourier transform of~$\{\beta_t\}$, denoted by $w_{\beta}(\lambda_j)$, and its corresponding periodogram $I_{\beta}(\lambda_j)$ at frequency $\lambda_j$ are defined as
\begin{equation*}
w_{\beta}(\lambda_j) = \frac{1}{(2\pi T)^{1/2}} \sum_{t=1}^T \beta_t \exp(it\lambda_j), \qquad I_{\beta}(\lambda_j) = \left| w_{\beta}(\lambda_j) \right|^2,
\end{equation*}
where $\exp$ denotes a characteristic function.

The LW approach is based on the likelihood of fitting a local approximation to the spectral density of~$\{\beta_t\}$ near frequency zero with a bandwidth of $J$. It assumes that
\begin{equation*}
f_{\mathrm{LW}}(\lambda;G,d) \sim G\lambda^{-2d} \quad \text{ as } \lambda \to 0+,
\end{equation*}
where $G>0$ is a scale parameter. 

Under the Gaussian frequency-domain approximation, the objective function of LW has the following form \citep[see][]{RM95}:
\begin{equation}\label{eq:2.2}
Q_{\mathrm{LW}}(G,d) = \frac{1}{J} \sum_{j=1}^{J}\left[\log\left(G\lambda_j^{-2d}\right)+\frac{I_{\beta}(\lambda_j)}{G\lambda_j^{-2d}}\right],
\end{equation}
where $\log(\cdot)$ denotes the natural logarithm, and $J$ is the total number of frequencies considered for Whittle-type estimation. The different variants of the LW estimator arise from alternative specifications of the spectral density near zero frequency.

For a fixed $d$, the scale parameter $G$ can first be concentrated by fixing $d$ and minimising the objective function with respect to $G$, and then substituted back into ~\eqref{eq:2.2}. This reduces the problem to an objective function of $d$ only. The resulting estimator of $G$ is given by
\begin{equation*}
\widehat{G}_{\mathrm{LW}}(d) = \frac{1}{J}\sum_{j=1}^{J} \lambda_j^{2d} I_{\beta}(\lambda_j).
\end{equation*}
Therefore, the LW estimator is obtained by minimising
\begin{equation*}
\widehat{d}_{\mathrm{LW}} = \arg\min_{d\in (-\frac{1}{2},\frac{1}{2})} \left[\log \widehat{G}_{\mathrm{LW}}(d) - \frac{2d}{J} \sum_{j=1}^J \log \lambda_j\right].
\end{equation*}

\subsection{Local Polynomial Whittle With Noise Estimator}\label{sec:2.3}

Proposed in \citet{FN08} and \citet{FNN12}, the LPWN estimator extends the LW estimators by allowing the short-memory component to vary smoothly near zero frequency. Motivated by the local polynomial bias-correction idea, we consider the working model
\begin{equation}
f_{\mathrm{LPWN}}(\lambda;G,d,\boldsymbol{\theta}) \sim G\lambda^{-2d}
\left\{1+h(d,\boldsymbol{\theta},\lambda)\right\},\label{eq:LPWN}
\end{equation}
where $h(d,\boldsymbol{\theta},\lambda) = \sum_{\ell=1}^{r}\theta_{\ell}\lambda^{2\ell} + \theta_{r+1}\lambda^{2d}$, and $\boldsymbol{\theta} = (\theta_1,\dots,\theta_r,\theta_{r+1})^{\top}$
denotes the vector of parameters, with $(\theta_{1},\dots,\theta_{r})$ corresponding to the polynomial correction terms in the short-memory component of the signal, and $\theta_{r+1}$ corresponding to the additive perturbation (noise) component. The polynomial terms $\sum_{\ell=1}^{r}\theta_{\ell}\lambda^{2\ell}$ account for the smooth, short-memory curvature near zero frequency, and $r$ is the total number of polynomial orders that account for bias correction; customarily $r = 1$ or 2. Additionally, the term $\theta_{r+1}\lambda^{2d}$ corresponds to the additive noise component. For the LPWN estimator, we will use the notation LPWN$_r$ to further denote the number of polynomial terms for bias correction. By plugging $h(d,\boldsymbol{\theta},\lambda)$ into~\eqref{eq:LPWN}, we obtain 
\begin{align*}
f_{\mathrm{LPWN}}(\lambda;G,d,\boldsymbol{\theta}) &\sim G\lambda^{-2d}\{1 + \sum_{\ell=1}^{r}\theta_{\ell}\lambda^{2\ell} + \theta_{r+1}\lambda^{2d}\} \\
&= G\lambda^{-2d}\Big(1+\sum_{\ell=1}^{r}\theta_{\ell}\lambda^{2\ell}\Big) + G\theta_{r+1}.
\end{align*}

The corresponding Whittle objective is defined as
\begin{equation*}
Q_{\mathrm{LPWN}}(G,d,\boldsymbol{\theta}) = \frac{1}{J} \sum_{j=1}^{J}
\left[\log\left[ G \lambda^{-2d} \left\{1+h(d,\boldsymbol{\theta},\lambda)\right\}\right] + \frac{I_{\beta}(\lambda_j)}{\left[ G \lambda^{-2d}
\left\{1+h(d,\boldsymbol{\theta},\lambda)\right\}\right]}
\right].
\end{equation*}

After concentration, the objective becomes
\begin{equation*}
\widetilde{Q}_{\mathrm{LPWN}}(d,\boldsymbol{\theta}) = \log \widehat{G}_{\mathrm{LPWN}}(d,\boldsymbol{\theta}) + \frac{1}{J}\sum_{j=1}^J \log\left\{\lambda_j^{-2d} \left(1+h(d,\boldsymbol{\theta},\lambda_j)\right)
\right\},
\end{equation*}
where
\begin{equation*}
\widehat{G}_{\mathrm{LPWN}}(d,\boldsymbol{\theta}) = \frac{1}{J}\sum_{j=1}^J
\frac{\lambda_j^{2d}I_{\beta}(\lambda_j)}{1+h(d,\boldsymbol{\theta},\lambda_j)}.
\end{equation*}

Therefore, the LPWN estimator is defined as
\begin{equation*}
\left(\widehat{d}_{\mathrm{LPWN}},\widehat{\boldsymbol{\theta}}\right)
= \arg\min_{d\in (0,1),\boldsymbol{\theta}\in R^{r+1}} \widetilde{Q}_{\mathrm{LPWN}}(d,\boldsymbol{\theta}).
\end{equation*}

Thus, these two estimators differ mainly in the local spectral approximation used near zero frequency. Since the focus of this study is on long-memory FARMA processes, where both long-memory dependence and short-memory FARMA dynamics may be present, we use the LPWN estimator in the subsequent analysis.

It is worth mentioning that the above estimators are only consistent within the range $d \in (0,1)$ and converge to the upper boundary when the true memory parameter exceeds 1. Therefore, for the cases where $d>1$, we apply a recursive integer-order differencing discussed in \citet{LRS23} in Algorithm~1.

\begin{algorithm}
\caption{Recursive Integer-Order Differencing for Memory Parameter Estimation}
\label{alg:1}
\begin{algorithmic}[1]
\Require Stationary or nonstationary univariate time series of scores $\{\beta_t\}_{t=1}^T$, small tuning parameter, $\gamma > 0$, customarily we use $\gamma=1/\log\{\lfloor 1+T^{0.65}\rfloor\}$.
\Ensure Final memory parameter estimate $\widehat d_{\mathrm{final}}$
\State Set $k=0$ and $\beta_t^{(0)}=\beta_t$.
\State Apply the selected estimator, such as the LPWN estimator, to $\{\beta_t^{(0)}\}$ and obtain $\widehat d(0)$.
\While{$\widehat d(k) \geq 1-\gamma$}
\State Set $k=k+1$.
\State Take the first difference of the previous series:
\begin{equation*}
\beta_{t}^{(k)} = \Delta \beta_t^{(k-1)} = \beta_t^{(k-1)}-\beta_{t-1}^{(k-1)}.
\end{equation*}
\State Apply the same selected estimator to $\{\beta_t^{(k)}\}$ and obtain $\widehat d(k)$.
\EndWhile
\State Calculate the final estimate as
\begin{equation*}
\widehat d_{\mathrm{final}} = \widehat d(k)+k.
\end{equation*}
\State \Return $\widehat d_{\mathrm{final}}$
\end{algorithmic}
\end{algorithm}

\section{Sieve bootstrap of stationary functional time series}\label{sec:3}

Let $L$ denote the backshift operator. With pre-filtering, we obtain a sequence of \textit{stationary} functional time series $\X_t(u)=(1-L)^{d}\Y_t(u)$, for $t=1,\dots, T$. Without loss of generality, we assume that the series has zero mean. The essence of the sieve bootstrap is to generate a functional time series of pseudo-random elements $\left\{\X_{t}^{*}(u)\right\}_{t=1}^{T}$ that adequately mimics the structure of the temporal dependence of the original functional time series $\left\{\X_t(u)\right\}_{t=1}^{T}$. Through Karhunen-Lo\`{e}ve expansion, random element $\X_t(u)$ can be decomposed into two parts:
\begin{align}
\X_t(u) = \sum^{\infty}_{k=1}\beta_{t,k}\phi_k(u) &= \sum^{K}_{k=1}\beta_{t,k}\phi_k(u)+\sum^{\infty}_{k=K+1}\beta_{t,k}\phi_k(u) \notag\\
	    &= \sum^{K}_{k=1}\beta_{t,k}\phi_k(u)+U_{t,K}, \label{eq:3.3}
\end{align}
where $(\phi_k(u), k = 1, 2,\dots, K)$ are the orthonormal eigenfunctions of the sample covariance operator, expressed as $\frac{1}{T}\sum^T_{t=1}\X_t\otimes \X_t$, $K$ represents the number of retained functional principal components, and $\bm{\beta}_t = (\beta_{t,1}, \beta_{t,2},\dots,\beta_{t, K})^{\top}$ denotes the associated principal component scores at time $t$. The first term is considered the main driving part of $\X_t(u)$, while the second term $U_{t, K}$ is white noise.

To select the retained number of components, we choose $K$ such that it explains at least 95\% of the total variance, expressed as
\begin{equation*}
K = \argmin\left\{\frac{\sum_{k=1}^K\lambda_k}{\sum_{k=1}^T\lambda_k}\geq 0.95\right\},
\end{equation*}
where $\lambda_k$ represents the $k$\textsuperscript{th} estimated eigenvalue from the functional principal component analysis.

With $K$ leading eigenfunctions $[\phi_1(u),\phi_2(u),\dots,\phi_K(u)]$, we project a time series of functions $\X_t(u)$ onto these orthonormal eigenfunctions and obtain a time series of estimated $K$-dimensional vectors of scores. These scores can then be modelled via
\begin{equation*}
\widehat{\bm{\beta}}_{t^{'}} = \sum^{p}_{\omega = 1}\widehat{\bm{A}}_{\omega,p}\widehat{\bm{\beta}}_{t^{'}-\omega} + \widehat{\bm{e}}_{t^{'}}, \qquad t^{'}=p+1,p+2,\dots,T,
\end{equation*}
where $\widehat{\bm{A}}_{\omega,p}$ denotes the VAR$(p)$ coefficients, and $\widehat{\bm{e}}_{t^{'}}$ are the estimated residuals. The order $p$ of the fitted VAR model is chosen using a corrected Akaike information criterion \citep{HT93}, that is, by minimising
\begin{equation*}
\text{AICC}(p) = T\log|\widehat{\bm{\Sigma}}_{e,p}|+\frac{T(TK+pK^2)}{T-K(p+1)-1},
\end{equation*}
over a range of values of $p$, where $\bm{\Sigma}=\frac{1}{T}\sum^T_{t^{'}=p+1}\widehat{\bm{e}}_{t^{'},p}\widehat{\bm{e}}_{t^{'},p}^{\top}$ is the variance of the residuals and $\widehat{\bm{e}}_{t^{'},p}$ are the residuals after fitting the VAR($p$) model to the $K$-dimensional time series of estimated principal component scores $(\widehat{\bm{\beta}}_1, \widehat{\bm{\beta}}_2, \dots, \widehat{\bm{\beta}}_T)$. The \verb|VARselect| function in the \textit{vars} package \citep{Pfaff08} in \Rlogo \ is used for selecting the optimal VAR order and estimating the parameters.

Since VAR is an autoregression, it requires a burn-in period to eliminate the effect of starting values. With a burn-in sample size of 100, we generate the vector pseudo-time series of scores
\begin{equation*}
\bm{\beta}_{t^{\diamond}}^{*} = \sum^p_{\omega=1}\widehat{\bm{A}}_{\omega,p}\bm{\beta}_{t^{\diamond}-\omega}^*+\bm{e}_{t^{\diamond}}^{*}, \quad t^{\diamond}=1,2,\dots,(T+100),
\end{equation*}
where we use the starting value $\bm{\beta}_{t^{\diamond}}^{*} = \bm{\widehat{\beta}}_t$ for $t^{\diamond}=1,2,\dots,p$. The bootstrap residuals $\bm{e}_{t^{\diamond}}^{*}$ are independent and identically distributed (i.i.d.) resampled from the set of centered residuals $\{\widehat{\bm{e}}_{t^{'}} - \overline{\bm{e}}_T\}$, and $\overline{\bm{e}}_T = \frac{1}{T-p}\sum^{T}_{t^{'}=p+1}\widehat{\bm{e}}_{t^{'}}$.

Based on the Karhunen-Lo\`{e}ve expansion in~\eqref{eq:3.3}, we obtain bootstrap curves
\begin{equation*}
\X_{t^{\diamond}}^{*}(u) = \sum^{K}_{k=1}\beta_{t^{\diamond},k}^*\widehat{\phi}_k(u)+U_{t^{\diamond},K}^{*}(u),
\end{equation*}
where $U_{t^{\diamond},K}^*$ are i.i.d. resampled from the set $\left\{\widehat{U}_{t,K}(u) - \overline{U}_T(u), t=1,2,\dots,T\right\}$, $\overline{U}_T(u) = \frac{1}{T}\sum^T_{t=1}\widehat{U}_{t,K}(u)$, and $\widehat{U}_{t,K}(u) = \X_t(u) - \sum^K_{k=1}\widehat{\beta}_{t,k}\phi_k(u)$. We discard the first 100 generated $\X_{t^{\diamond}}^{*}(u)$ observations and keep $[\X_{101}^*(u),\X_{102}^*(u),\dots,\X_{T+100}(u)]$ as the bootstrap functional pseudo time series.

In Algorithm~2, we present a workflow for computing sieve bootstrap samples for stationary functional time series \citep[see also][]{Paparoditis18}.
\begin{algorithm}
\caption{Sieve Bootstrap for Stationary Functional Time Series}
\label{alg:2}
\begin{algorithmic}[1]
\Require Centered functional time series $\{\mathcal{X}_t(u)\}_{t=1}^T$, number of principal components $K$, VAR model order $p$, burn-in period of $100$
\Ensure Bootstrap functional time series $\{\mathcal{X}_{t^{\diamond}}^{*}(u)\}_{t^{\diamond}=1}^{T+100}$, where $^{*}$ denotes one bootstrap replication
\State Estimate the first $K$ number of eigenvalues $\widehat{\lambda}_1>\widehat{\lambda}_2>\cdots>\widehat{\lambda}_K$ and the corresponding eigenfunctions $\left\{\widehat{\phi}_k(u)\right\}_{k=1}^K$ from sample covariance operator $\widehat{C}_{\mathcal{X}}=T^{-1}\sum_{t=1}^T \mathcal{X}_t \otimes \mathcal{X}_t$. 
\State Calculate $K$-dimensional Fourier coefficients $\widehat{\boldsymbol{\beta}}_t = \left(\widehat{\beta}_{t,k}=\langle \mathcal{X}_t, \widehat{\phi}_k \rangle,k=1,2,\ldots,K\right)^\top$ and the functional residuals $\widehat{U}_{t,K}(u) = \mathcal{X}_t(u) - \sum_{k=1}^K \widehat{\beta}_{t,k}\widehat{\phi}_k(u)$.
\State Fit VAR($p$) model to the $K$-dimensional time series $\left\{\widehat{\boldsymbol{\beta}}_t\right\}_{t=1}^T$ as $\widehat{\boldsymbol{\beta}}_t = \sum_{\omega=1}^p \widehat{\mathbf{A}}_{\omega,p} \widehat{\boldsymbol{\beta}}_{t-\omega} + \widehat{\boldsymbol{e}}_t$.
\State Center VAR residuals as $\widetilde{\boldsymbol{e}}_t = \widehat{\boldsymbol{e}}_t - \frac{1}{T-p}\sum_{j=p+1}^T \widehat{\boldsymbol{e}}_j$, and functional residuals as $\widetilde{U}_{t,K}(u) = \widehat{U}_{t,K}(u) - \frac{1}{T}\sum_{j=1}^T \widehat{U}_{j,K}(u)$. Initialize $\boldsymbol{\beta}_t^* = \widehat{\boldsymbol{\beta}}_t$ for $t=1, \dots, p$
\For{$t = p+1$ \textbf{to} $T+100$}
\State Sample $\boldsymbol{e}_{t^{\diamond}}^{*}$ i.i.d. from $\{\widetilde{\boldsymbol{e}}_j\}_{j=p+1}^T$ and calculate $\boldsymbol{\beta}_{t^{\diamond}}^{*}=(\beta_{1,t^{\diamond}}^{*},\beta_{2,t^{\diamond}}^{*},\ldots,\beta_{K,t^{\diamond}}^{*})^\top$ as $\boldsymbol{\beta}_{t^{\diamond}}^{*} = \sum_{\omega=1}^p \widehat{\mathbf{A}}_{\omega,p} \boldsymbol{\beta}_{t^{\diamond}-\omega}^* + \boldsymbol{e}_{t^{\diamond}}^{*}$. 
\State Sample $U_{t^{\diamond},K}^*(u)$ i.i.d. from $\{\widetilde{U}_{j,K}(u)\}_{j=1}^T$ and calculate $\mathcal{X}_{t^{\diamond}}^{*}(u)=\sum_{k=1}^K \beta_{t^{\diamond},k}^*\phi_k(u) + U_{t^{\diamond},K}^*(u)$
\EndFor
\State \Return $\{\mathcal{X}_{101}^*(u),\mathcal{X}_{102}^*(u), \dots, \mathcal{X}_{T+100}^*(u)\}$
\end{algorithmic}
\end{algorithm}

\section{Bias-correction method via the prefiltered sieve bootstrapping}\label{sec:4}

For long-range dependent functional data, we begin by applying the LPWN estimator to estimate an initial memory parameter $\widehat{d}$. With the estimated $\widehat{d}$, we form the prefiltered series via fractional differencing, converting long-range dependent series $\Y_t(u)$ into a stationary short-range dependent series $\X_t(u)$. Using the functional sieve bootstrap method in Section~\ref{sec:3}, we generate $B=399$ bootstrap samples~$\X_{t}^{*}(u)$ \citep[see][on the number of bootstraps]{DM00}. Applying fractional integration, we obtain bootstrap samples~$\Y_{t}^{*}(u)$. For each $\Y_{t}^{*}(u)$, we apply the LPWN estimator to obtain a distribution of the memory parameter~$d$. From the bootstrap $\widehat{d}^*$, we can estimate the bias and obtain confidence intervals for $d$. In Algorithm~3, we show a workflow for implementing our bias-correction method via the prefiltered sieve bootstrap.
\begin{algorithm}[!htb]
\caption{Prefiltered sieve bootstrap for LPWN estimation and bias correction}\label{alg:3}
\begin{algorithmic}[1]
\Require Observed series $\{\Y_t\}_{t=1}^T$, number of bootstraps $B$, (optional) sieve order selection rule
\Ensure Bias-corrected long-memory estimate $\widehat d_{\mathrm{BC}}$
%\vspace{0.25em}
\State \textbf{Initial semiparametric estimation (LPWN) for a chosen polynomial order $r$, customarily $r=1$ or $2$.}
\State Compute $\widehat d \gets \textsc{LPWN}(\{\Y_t\}_{t=1}^n)$.
%\vspace{0.25em}
\State \textbf{Key step: prefiltering.}
\State Let $L$ denote the backshift operator; form the prefiltered series (fractional differencing)
\begin{equation*}
\X_t \gets (1-L)^{\widehat d} \Y_t,\qquad t=1,2,\dots,T.
\end{equation*}
\State \textbf{Functional sieve bootstrap on prefiltered data.}
\State Fit an autoregressive (AR) sieve model to $\{\X_t\}$ (order chosen by rule), obtaining fitted coefficients and residuals.
\For{$b=1,\dots,B$}
\State Generate a bootstrap series $\X_t^{*}$ using the functional sieve bootstrap based on the fitted AR sieve for $\X_t$.
\vspace{0.25em}
\State \textbf{Re-inject long memory.}
\State Apply the inverse fractional filter to obtain bootstrap samples on the original scale:
\begin{equation*}
\Y_t^{*} \gets (1-L)^{-\widehat d} \X_t^{*}.
\end{equation*}
\State \textbf{Re-estimate $d$ on bootstrap sample.}
\State Compute $\widehat d^{(b)} \gets \textsc{LPWN}(\{\Y_t^{*}\}_{t=1}^{T})$.
\EndFor
\vspace{0.25em}
\State \textbf{Bootstrap bias estimate and correction.}
\State Compute the bootstrap mean, $\overline d^{(b)} \gets \frac{1}{B}\sum_{b=1}^B \widehat d^{(b)}$, and estimate bias, $\widehat{\mathrm{bias}} \gets \overline d^{(b)} - \widehat d$.
\State Bias-corrected estimator:
\begin{equation*}
\widehat d_{\mathrm{BC}} \gets \widehat d - \widehat{\mathrm{bias}} \;=\; 2\widehat d - \overline d^{(b)} .
\end{equation*}
\State \Return $\widehat d_{\mathrm{BC}}$
\end{algorithmic}
\end{algorithm}

\section{Monte Carlo Simulation study}\label{sec:5}

\subsection{Simulation data generating process setting}\label{sec:5.1}

We simulate $\{\Y_{t}(u)\}_{t=1}^{T}$ from a functional FARFIMA$(p,d,q)$ model on $\mathcal{C} = [0,1]$, defined by
\begin{equation*}
\Y_t(u) = (1 - L)^{-d}\X_t(u),
\end{equation*}
where $\{\X_t\}$ is a short-memory process following a functional FARMA$(p,q)$ model given by
\begin{equation*}
\X_{t}(u) - \sum_{i=1}^{p} \int_{0}^{1} \phi_i(u,v) \X_{t-i}(v)\, dv
= \eta_t(u) + \sum_{j=1}^{q} \int_{0}^{1} \theta_j(u,v) \eta_{t-j}(v)\, dv,
\end{equation*}
and $\{\eta_t\}$ are i.i.d.\ standard Brownian motions on $[0,1]$.

For the simulation of $\{\X_t(u)\}_{t=1}^{T}$, we consider the following three general cases:
\begin{description}
    \item[Case~1]  $p = 1,\; q = 0,\quad 
\phi_1(u,v) = c_1 \exp\!\left\{ -\frac{u^2 + v^2}{2} \right\}$;
\item[Case~2] $p=0, \; q=1$,\quad $\theta_1(u,v) = c_2 \min(u,v)$;
\item[Case~3] $p = 1,\; q = 1,\quad 
\phi_1(u,v) = c_1 \exp\!\left\{ -\frac{u^2 + v^2}{2} \right\},\quad 
\theta_1(u,v) = c_2 \min(u,v)$.
\end{description}

The three cases correspond to FARFIMA(1, $d$, 0), FARFIMA(0, $d$, 1) and FARFIMA(1, $d$, 1) respectively, and the constants $c_1$ and $c_2$ are chosen such that the corresponding integral operators satisfy $\psi = \|\phi_1\|$ and $\nu = \|\theta_1\| \in \{0.2, 0.5, 0.8\}$, representing weak, moderate, and strong short-range dependence, respectively.

Moreover, we consider sample sizes $T \in \{500, 1000, 2000\}$ and memory parameters $d \in \{0.1, 0.2, 0.3$, $0.4, 0.5, 0.6, 0.8, 1.0, 1.2, 1.4\}$. The range of $d$ covers both stationary ($d < 0.5$) and non-stationary regimes ($d \geq 0.5$), including strongly non-stationary cases ($d > 1$). For each configuration, 100 independent replications are generated with pseudorandom seeds.

To evaluate the performance of the proposed methods, we consider the LW and DFA as benchmark estimators, which are recommended in \citet{Shang22}. The proposed LPWN estimators, both LPWN$_1$ and LPWN$_{2}$, are used to assess their ability to reduce bias induced by short-run dependence. Their bootstrap-corrected versions, denoted BC-LPWN$_1$ and BC-LPWN$_{2}$, are considered to examine the effectiveness of bootstrap adjustment on the LPWN estimator, which is a bias-corrected estimator.

\subsection{Simulation estimation accuracy}\label{sec:5.2}

\subsubsection{FARFIMA\texorpdfstring{$(1,d,0)$}{(1,d,0)} Setting}

We begin with the FARFIMA$(1,d,0)$ setting, assessing and comparing estimation accuracy in terms of average bias and mean squared error (MSE), computed over 100 replications for each configuration, that is, the sample size and the strength of short-run dependence.

\begin{figure}[!htb]
\centering
\includegraphics[width=0.67\textwidth]{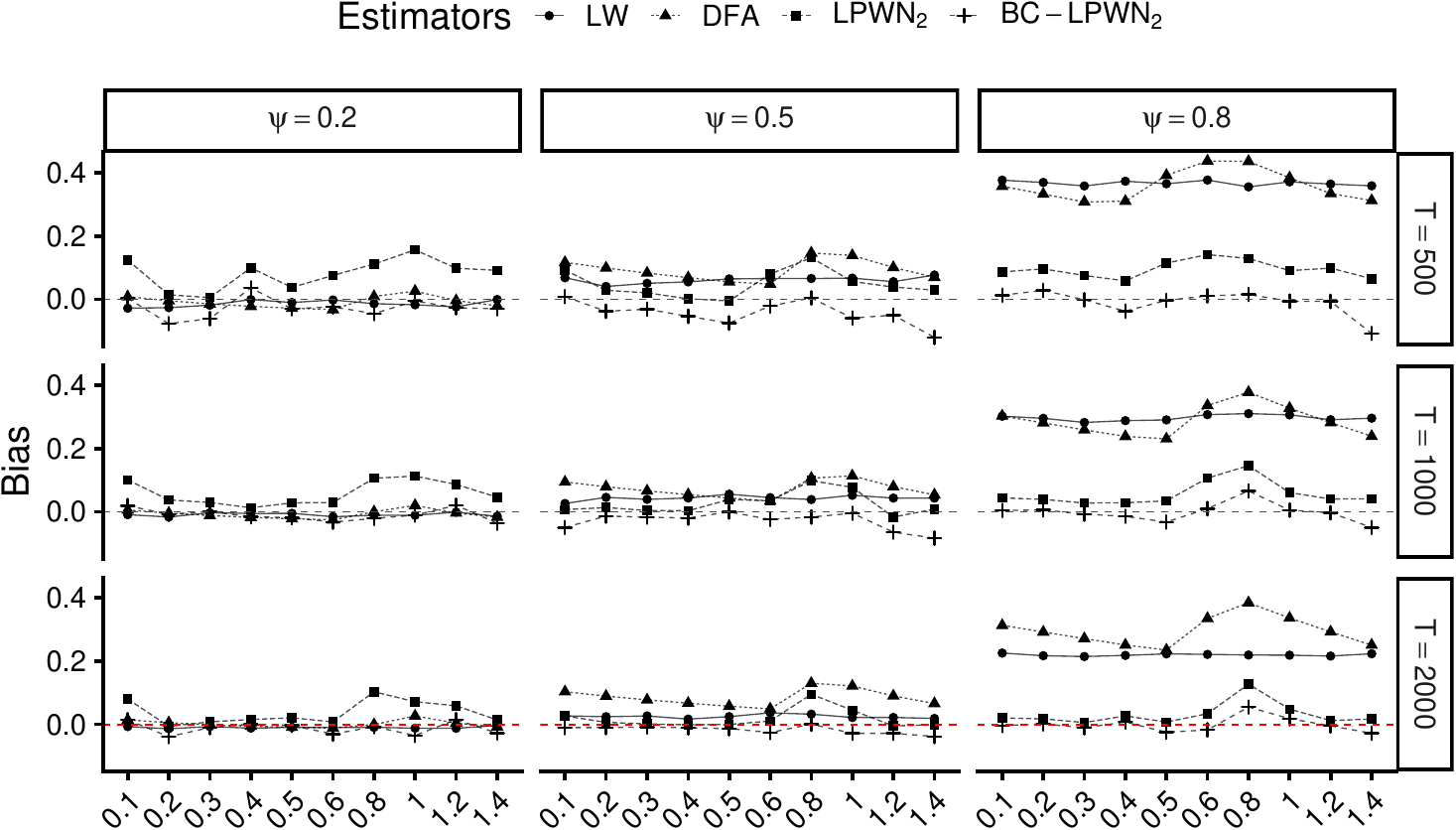} \\
\vspace{.05in}
\includegraphics[width=0.67\textwidth]{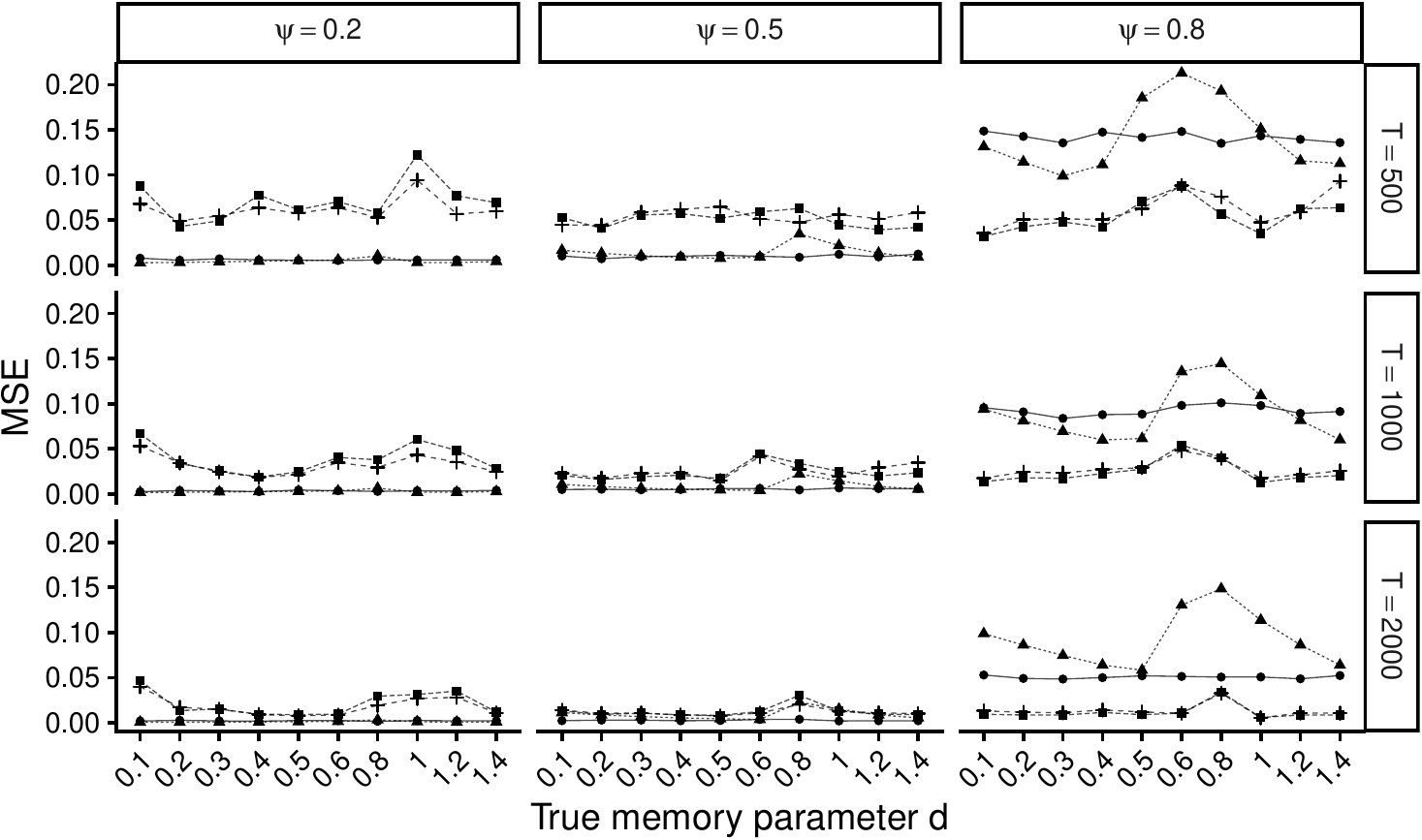}
\caption{\small As measured by the bias and MSE, finite-sample accuracy comparison of LW, DFA, LPWN$_{2}$, and BC-LPWN$_{2}$ for the FARFIMA$(1,d,0)$ setting across different values of $d$, $T$, and $\psi$.}
\label{fig:1}
\end{figure}

In Figure~\ref{fig:1}, the benchmark estimators LW and DFA perform well when the $\psi$ is weak to moderate ($\psi = \{0.2, 0.5\}$), with small bias and MSE across all values of $d$. However, when the $\psi$ becomes strong ($\psi = 0.8$), both estimators exhibit substantially increased bias and MSE, indicating the inflated estimation of $d$ due to strong short-run dependence. As the sample size increases, both bias and MSE decrease across all methods, indicating improved estimation accuracy.

The proposed LPWN$_{2}$ estimator reduces bias relative to LW, and the bootstrap-corrected version BC-LPWN$_{2}$ further improves bias, often bringing it close to zero. Additionally, results for the first-order correction, LPWN$_1$ and BC-LPWN$_1$, are reported in Figures~\ref{fig:sup1}, \ref{fig:sup2} and~\ref{fig:sup3} in the Appendix and exhibit similar patterns, although the second-order correction consistently achieves greater bias reduction. However, this improvement comes at the cost of increased variability, resulting in larger MSE compared to LW and DFA, particularly under weak or moderate autoregressive strength~$\psi$. When sample size increases from $T=500$ to 2000, the MSE values are about the same for all four estimators. This reflects a clear bias-variance trade-off under different configurations of data. Note that when $d$ is close to 1, all estimators exhibit noticeable instability, leading to spikes in bias. This behaviour is likely driven by the iterative estimation procedure mentioned in Algorithm~1 which becomes less stable near the unit-root boundary.

\subsubsection{FARFIMA\texorpdfstring{$(0,d,1)$}{(1,d,1)} Setting}\label{sec:5.2.2}

\begin{figure}[!htb]
\centering
\includegraphics[width=0.67\textwidth]{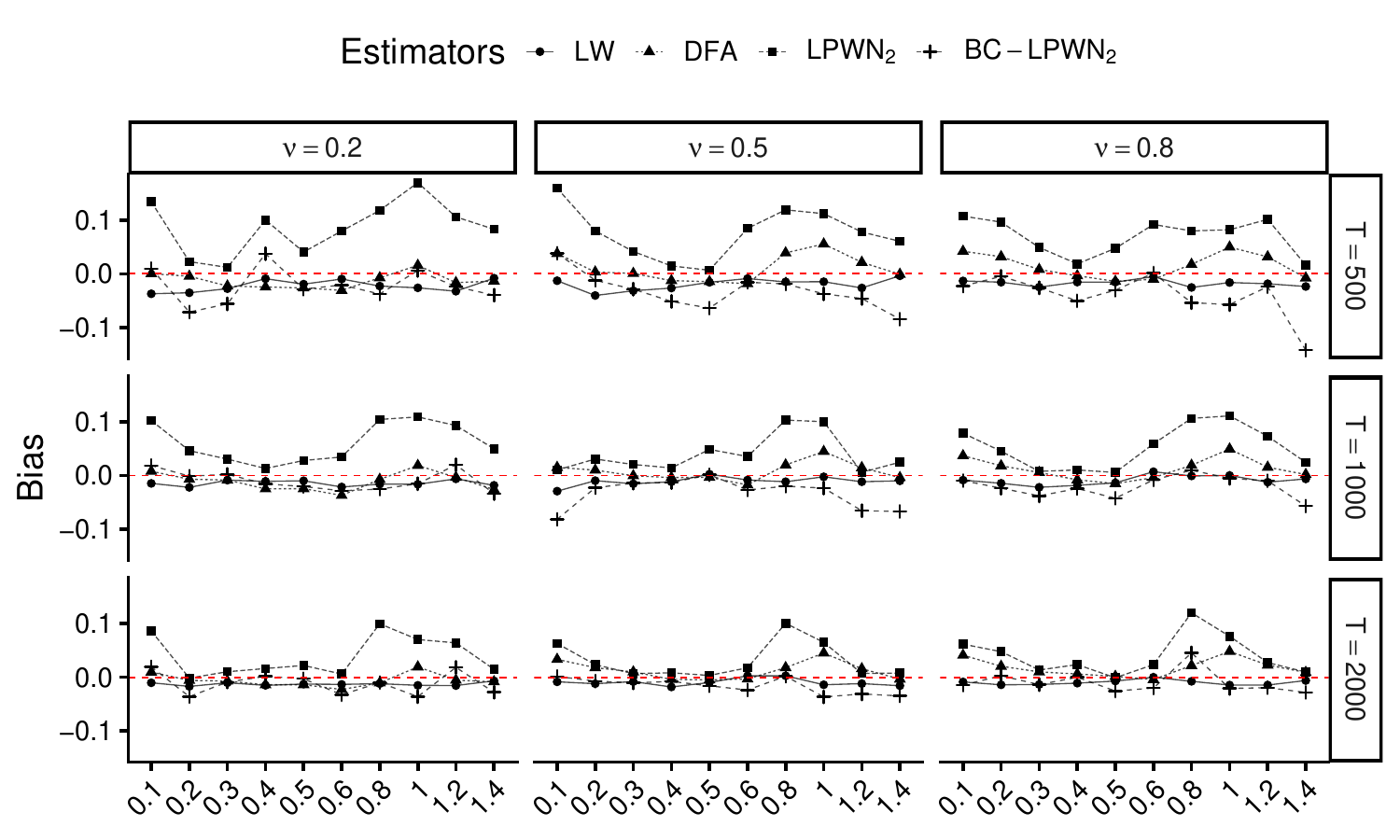} \\
\vspace{.05in}
\includegraphics[width=0.67\textwidth]{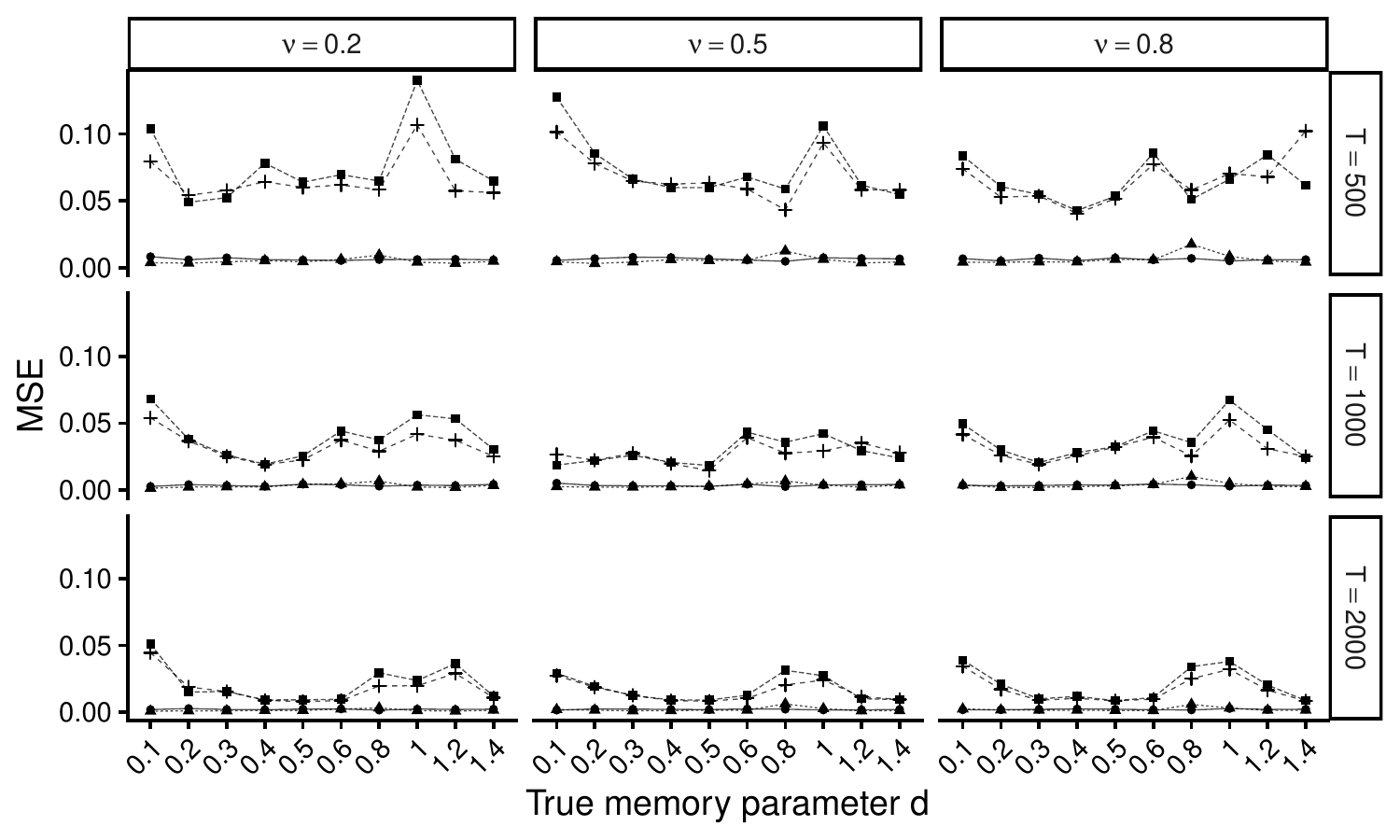}
\caption{\small As measured by the bias and MSE, finite-sample accuracy comparison of LW, DFA, LPWN$_{2}$, and BC-LPWN$_{2}$ for the FARFIMA$(0,d,1)$ setting across different values of $d$, $T$, and $\nu$.}
\label{fig:2}
\end{figure}

We also consider the FARFIMA$(0,d,1)$ setting to examine the influence of the moving-average component on memory estimation. The results in Figure~\ref{fig:2} show that varying the MA strength has only a limited effect on estimation accuracy as all estimation methods achieve low bias overall, and both the bias and MSE remain broadly similar across different MA strength. This suggests that estimation performance is driven primarily by autoregressive dependence rather than moving-average dependence. Overall, all estimators exhibit relatively small bias across the full range of $d$, with LW and BC-LPWN$_2$ consistently producing the smallest bias. As in the FARFIMA$(1,d,0)$ setting, increasing the sample size improves the estimation accuracy of all methods.

The MSE results follow a pattern similar to that observed in the FARFIMA$(1,d,0)$ setting. LW and DFA generally attain the smallest MSE across the configurations, whereas LPWN$_2$ and BC-LPWN$_2$ exhibit a larger MSE due to their higher estimation variability. Overall, the inclusion of the moving-average component does not substantially alter the relative performance of the estimators.

\subsubsection{FARFIMA\texorpdfstring{$(1,d,1)$}{(1,d,1)} Setting}\label{sec:5.2.3}

We last consider the FARFIMA$(1,d,1)$ setting, which incorporates both autoregressive and moving-average dependence. Building on the FARFIMA$(0,d,1)$ results, this setting allows us to examine the interaction effects of the moving-average component when it is combined with an autoregressive term on the estimation of the memory parameter. We therefore first evaluate estimation performance across different combinations of $\psi$ and $\nu$ at $T=2000$ to compare the relative effects of the AR and MA components.

\begin{figure}[!htb]
\centering
\includegraphics[width=0.67\textwidth]{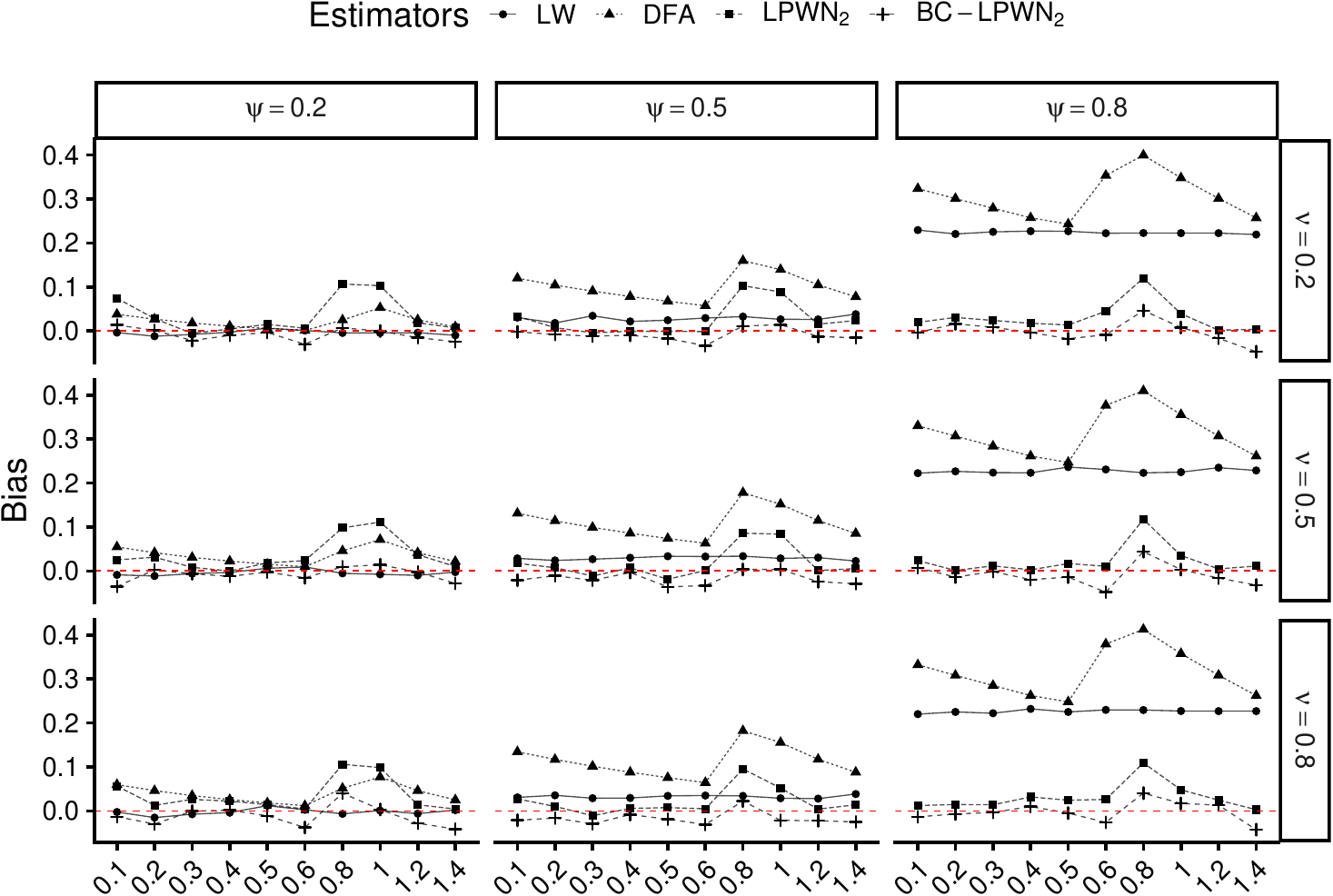} \\
\vspace{.2in}
\includegraphics[width=0.67\textwidth]{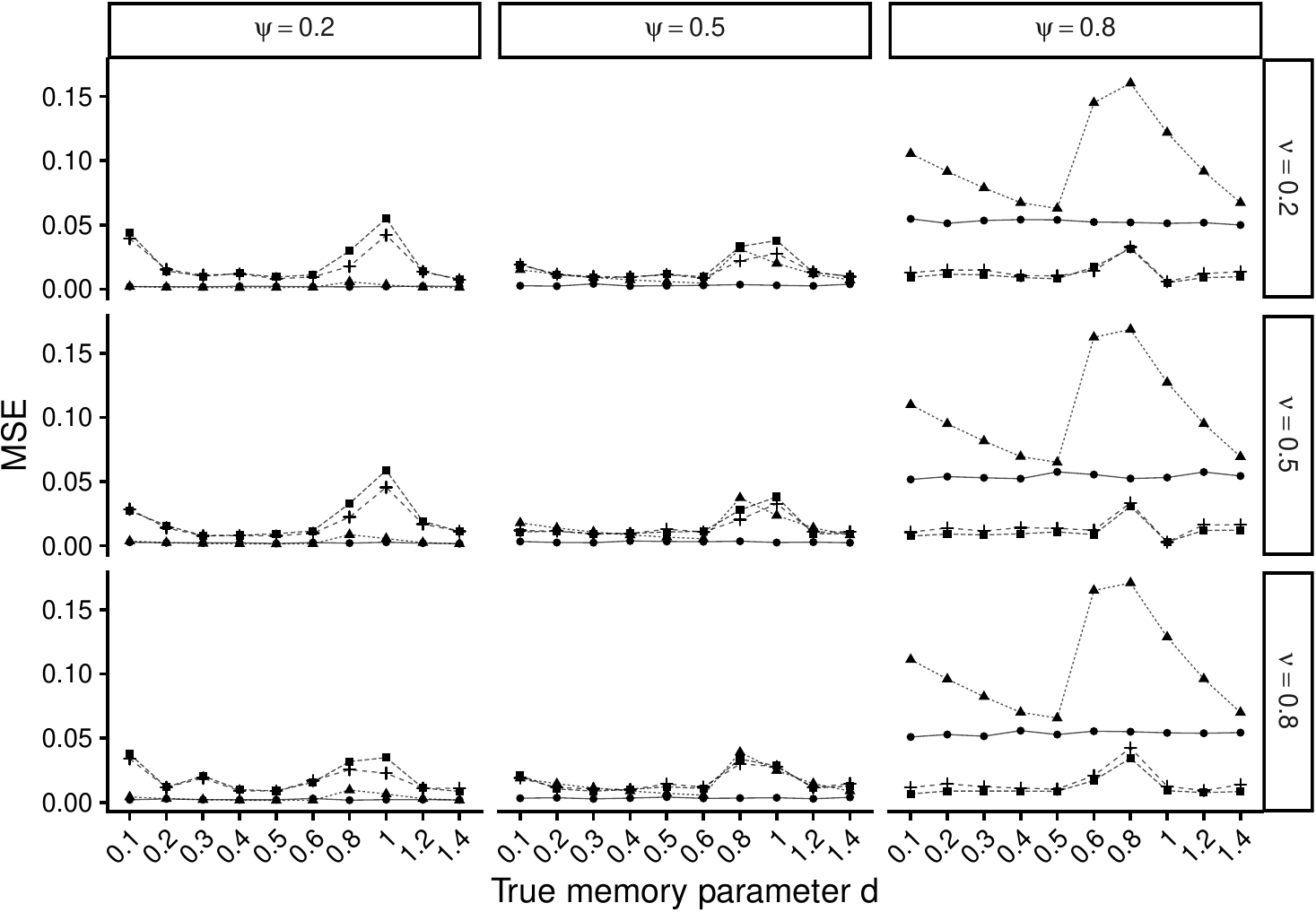}
\caption{\small As measured by the bias and MSE, finite-sample accuracy comparison of LW, DFA, LPWN$_{2}$, and BC-LPWN$_{2}$ for the FARFIMA$(1,d,1)$ setting across different values of $d$, $\psi$ and $\nu$ under $T = 2000$.}
\label{fig:3}
\end{figure}

In Figure~\ref{fig:3}, the estimation accuracy is largely unaffected by $\nu$ across all methods. In contrast, the autoregressive component continues to play a dominant role, with stronger dependence leading to increased bias and MSE. This indicates that short-run dependence arising from the autoregressive component has a much stronger impact on long-memory estimation than that induced by the moving average component.

\begin{figure}[!htb]
\centering
\includegraphics[width=0.7\textwidth]{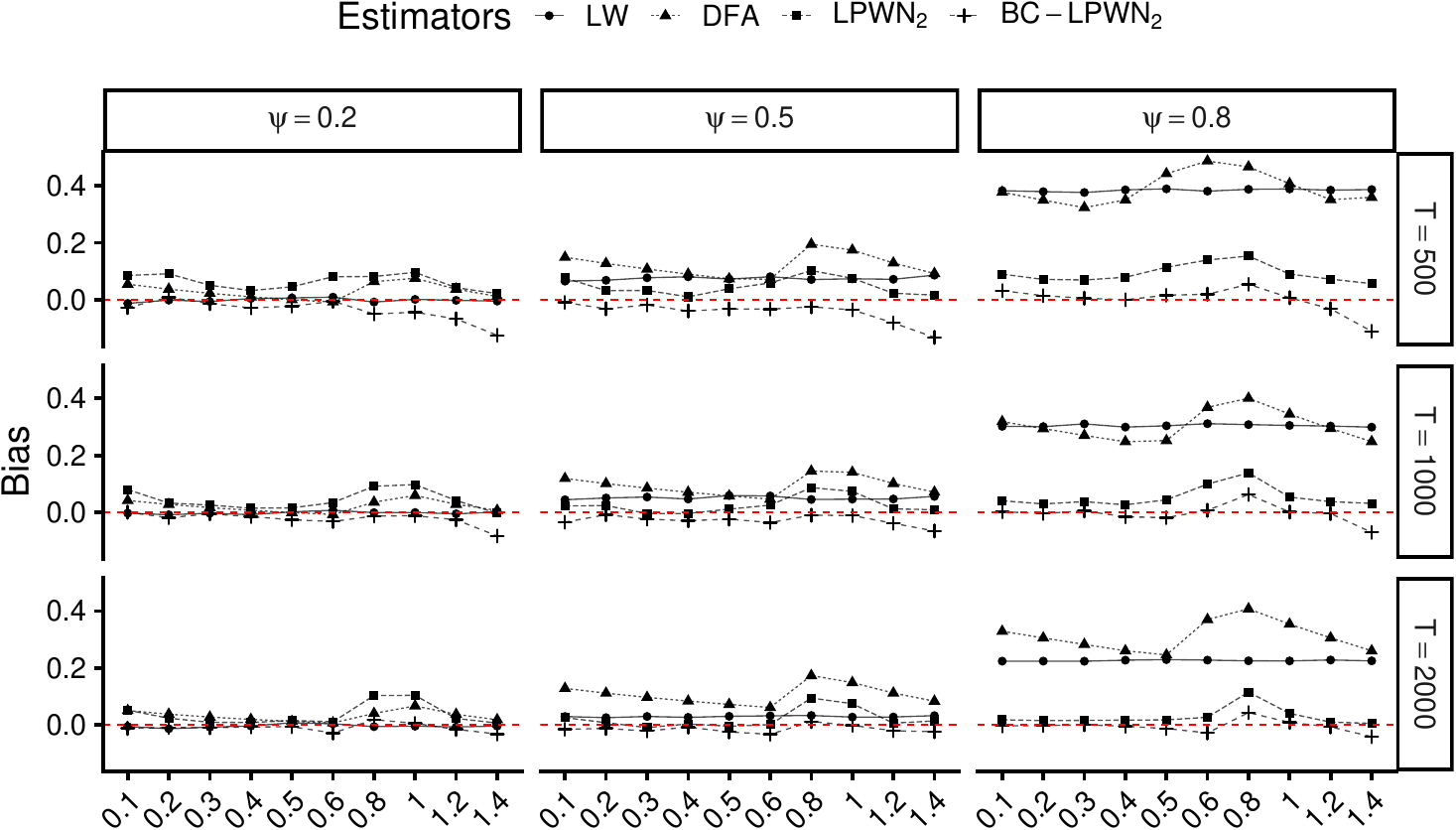}\\
\vspace{.2in}
\includegraphics[width=0.7\textwidth]{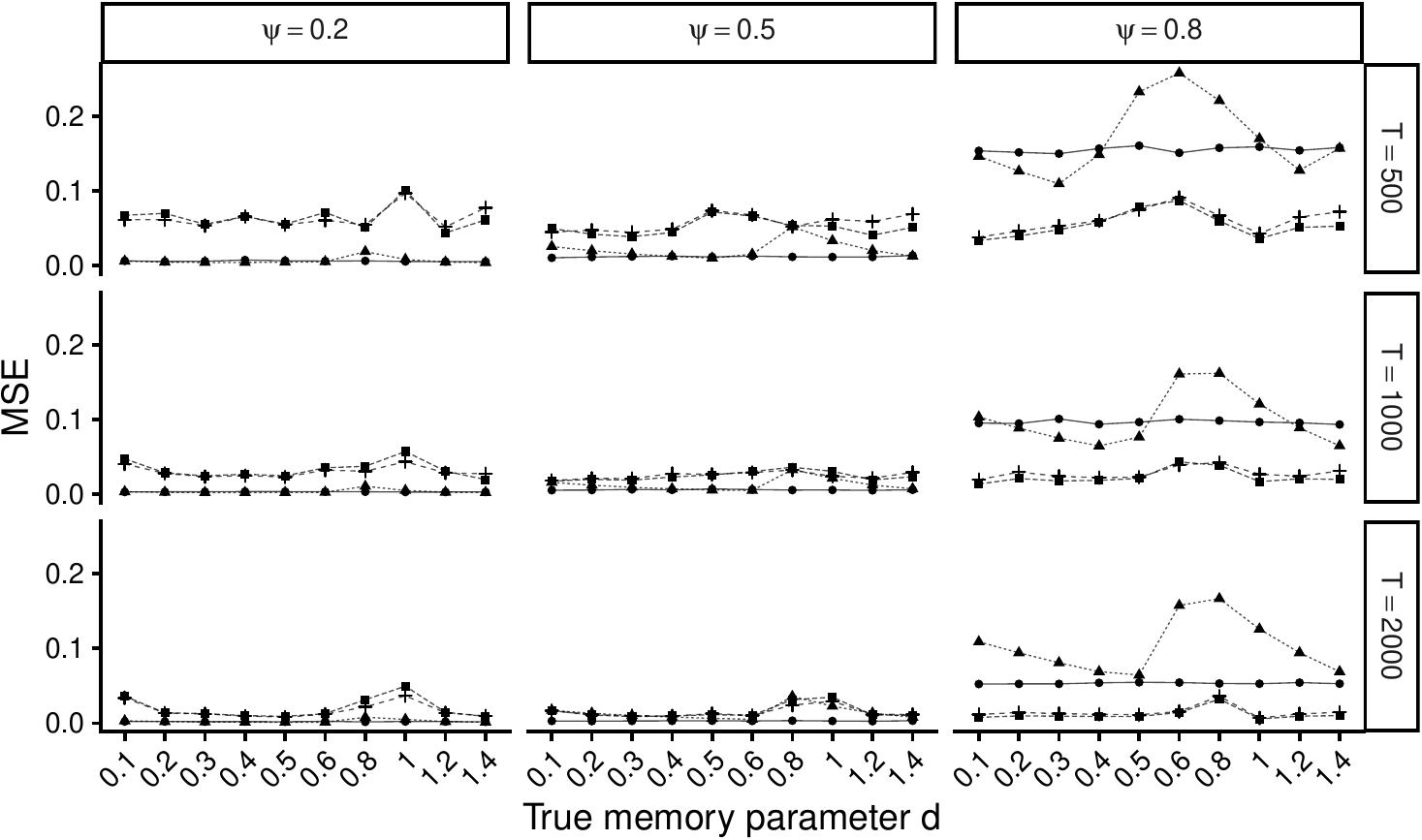}
\caption{\small Finite-sample performance comparison of LW, DFA, LPWN$_2$, and BC-LPWN$_2$ for the $FARFIMA(1,d,1)$ setting across different values of $d$, $T$, and $\psi$.}
\label{fig:4}
\end{figure}

Motivated by this observation, we further average the estimation accuracy over different $\nu$ and focus on the effects of $\psi$ and $T$. The results closely mirror those observed in the FARFIMA$(1,d,0)$ setting. In Figure~\ref{fig:4}, the LW and DFA perform well under weak dependence but under-perform significantly when the $\psi$ is large. Similarly, LPWN$_{2}$ reduces bias relative to LW, while BC-LPWN$_{2}$ further improves bias at the expense of increased MSE when $\psi$ is low or moderate. 

The detailed numerical results are also shown in Appendix~\ref{sec:S2}.

\section{Empirical data analysis}\label{sec:6}

We apply the proposed estimators to two empirical functional datasets: Swedish age-specific mortality rates and Canadian zero-coupon bond yield curves. Both datasets are regarded as exhibiting strong persistence and long-range dependence, making them appropriate applications for the proposed methodology. 

For each dataset, the memory parameter $d$ is estimated using the six competing estimators considered throughout this paper: LW, DFA, LPWN$_1$, LPWN$_2$, BC-LPWN$_1$, and BC-LPWN$_2$. To quantify estimation uncertainty, bootstrap confidence intervals are constructed using 399 bootstrap replications. The resulting bootstrap distributions and empirical confidence intervals are computed to assess the stability and uncertainty associated with each estimator.

\subsection{Swedish age-specific mortality rates}
\label{sec:6.1}

We first apply both benchmark estimation estimators and the proposed sieve-bootstrap bias-corrected estimators to Swedish age-specific mortality data obtained from the Human Mortality Database (\url{https://mortality.org/Country/Country?cntr=SWE}). In this dataset, we consider the annual age-specific death rates for both female and male populations, spanning the period from 1751 to 2024, yielding 274 annual observations. Based on existing literature in demography \citep{FSB25} and actuarial science \citep{PYC21}, mortality rates are commonly regarded as highly persistent functional time series and are often modelled as approximately integrated processes with a memory parameter close to one \citep[see, e.g.,][]{LRS23, SS26}. Therefore, this dataset provides a natural empirical application for investigating long-memory point estimation and bootstrapping uncertainty quantification in functional time series settings.

Following standard practice in mortality modelling, we analyse the natural logarithm of the mortality rates. Since some raw mortality rates are bounded above by one at higher ages, we truncate the ages from 0 to 89 in a single year of age, with mortality rates aged 90 and above combined into a single age group. In some years, zero values or missing mortality rates are then replaced by a small constant, $10^{-4}$, prior to the logarithmic transformation to avoid undefined values from $\log(0)$ \citep[see also][]{SS26}.

\begin{figure}[!htb]
\centering
\subfloat[Female data]
{\includegraphics[width=0.48\textwidth]{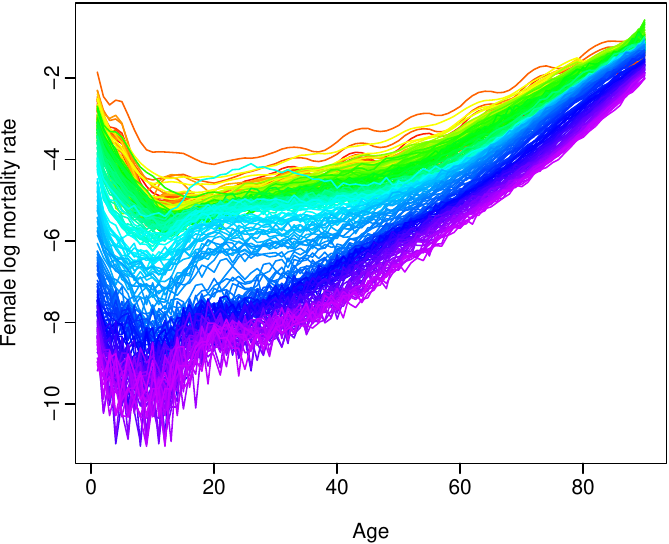}}
\quad
\subfloat[Male data]
{\includegraphics[width=0.48\textwidth]{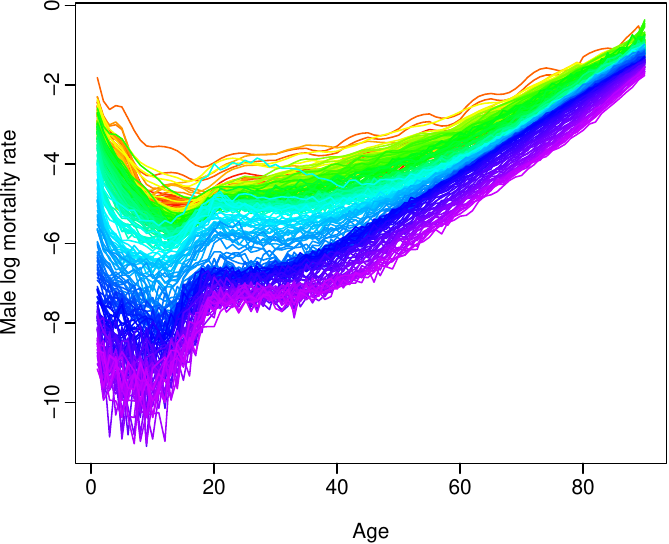}}
\caption{\small Rainbow plot of female and male annual age-specific log mortality rate from 1751 to 2024.}\label{fig:5}
\end{figure}

Figure~\ref{fig:5} presents rainbow plots of female and male log-mortality curves, respectively. The curves exhibit strong temporal persistence along with functional dependence across ages. In particular, mortality rates decrease during childhood and early adulthood before increasing steadily at older ages, while the gradual evolution across years suggests strong long-range dependence in the temporal dynamics. 

\begin{figure}[!htb]
\centering
\subfloat[Female data]
{\includegraphics[width=0.48\textwidth]{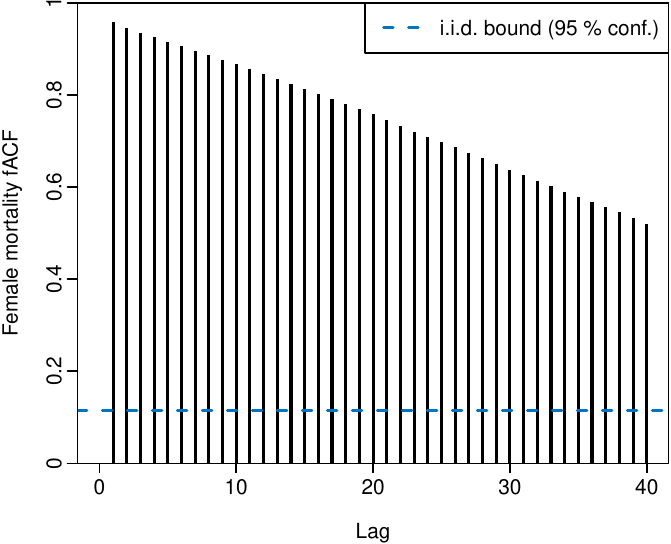}}
\quad
\subfloat[Male data]
{\includegraphics[width=0.48\textwidth]{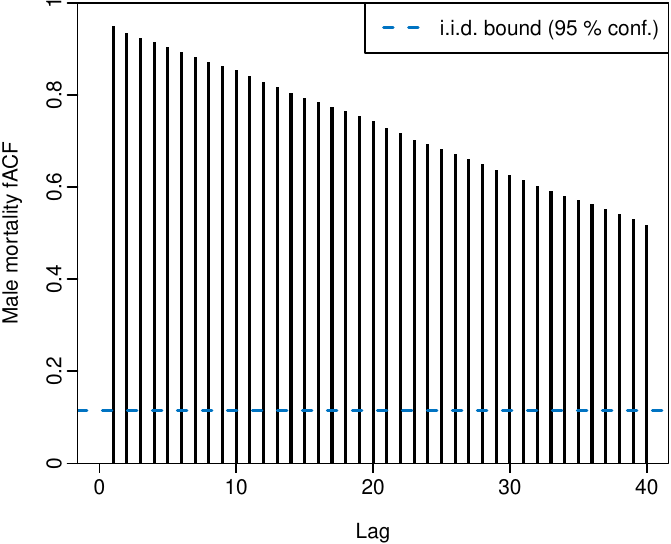}}
\caption{\small Functional autocorrelation plot of female and male annual age-specific log mortality rate from 1751 to 2024.}\label{fig:6}
\end{figure}

In Figure~\ref{fig:6}, we display functional autocorrelation plots of female and male log-mortality curves, respectively. Even at lag 40, the functional autocorrelation functions are still above the threshold line that indicates independence. 

\begin{figure}[!htb]
\centering
\includegraphics[width=0.65\textwidth]{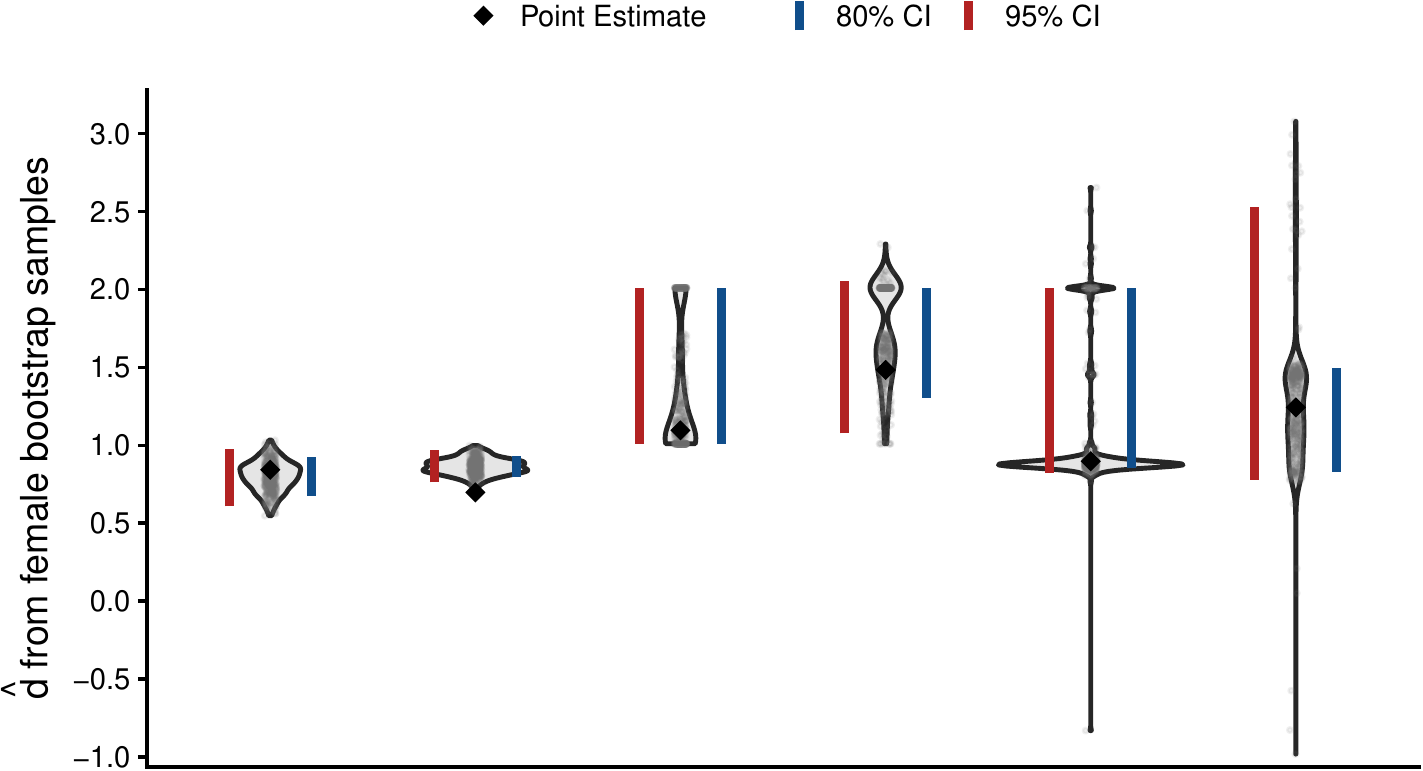} \\
\includegraphics[width=0.65\textwidth]{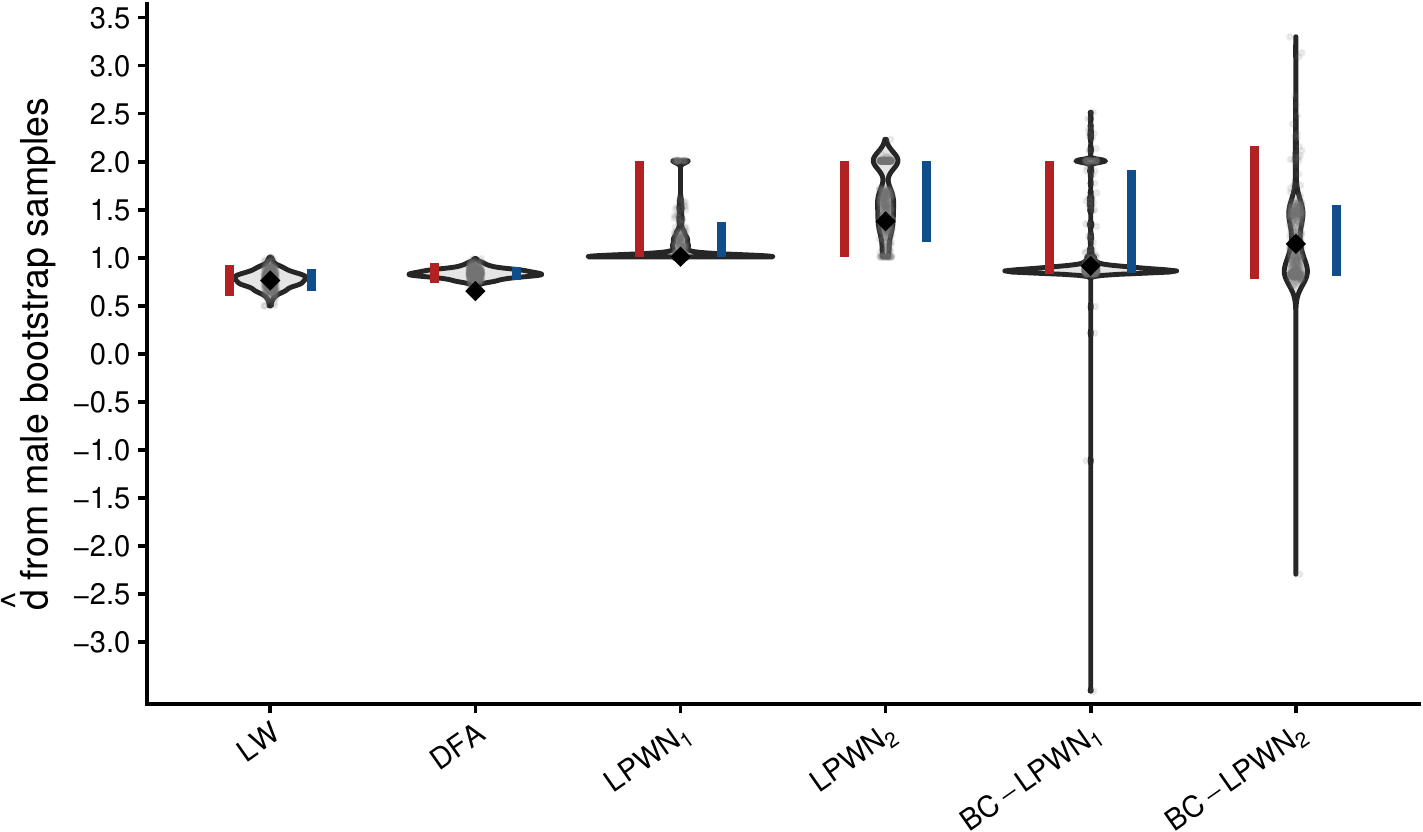}
\caption{\small Bootstrap distributions for estimated $d$, using 399 bootstrap samples, together with the associated empirical 80\% and 95\% confidence intervals across all estimators for the Swedish mortality data. The upper plot corresponds to female mortality memory estimation, while the lower plot corresponds to male mortality memory estimation.}
\label{fig:7}
\end{figure}

Figure~\ref{fig:7} shows the bootstrap distributions with the associated nominal 80\% and 95\% confidence intervals for each estimator. The estimated point memory parameters are near one across most estimators, supporting the claim that mortality rates behave approximately as integrated functional processes of order one. Additionally, the LW and DFA estimators produce relatively concentrated bootstrap distributions and narrower confidence intervals, whereas LPWN-based estimators exhibit larger variability and wider interval lengths, consistent with the findings in Section~\ref{sec:5.2}.

A noticeable feature of LPWN$_1$ and LPWN$_2$ is the accumulation of bootstrap estimates near values such as \(1.01\) and \(2.01\). This occurs because the LPWN optimization constrains the local memory parameter within \((0,1)\), while the iterative differencing procedure reconstructs the final estimate as $1+\widehat{d}$ or $2+\widehat{d}$. Consequently, many bootstrap replications accumulate near the boundary values. The sieve-bootstrap bias-corrected estimators BC-LPWN$_1$ and BC-LPWN$_2$ largely mitigate this boundary effect and produce smoother bootstrap distributions, although at the expense of increased variability due to the nested bootstrap procedure.

\subsection{Term-specific Canadian yield curve}\label{sec:6.2}

We next apply estimators to the end-of-day Canadian yield curves for the period spanning 4 July 2022 to 31 December 2025; the data are publicly available at \url{https://www.bankofcanada.ca/rates/interest-rates/bond-yield-curves/}. In the considered dataset, yields are observed daily (excluding non-trading days) on a fine grid of maturities: specifically, zero-coupon bond yields at 120 regularly spaced maturities from 0.25 to 30 years. We consider these data as functional data with a maturity continuum. The number of all available daily observations over the entire time span is 545. Similar yield curve data were previously analyzed by \citet{SS26}, although our study considers a longer observation period. The yield curve data are commonly regarded as highly persistent functional time series. Indeed, \citet{SS26} applied the sequential \(V_0\)- and \(V_1\)-tests and concluded that the yield curve process behaves as an \(I(1)\) functional time series at the 5\% significance level. 

\begin{figure}[!htb]
\centering
\subfloat[Rainbow plot]
{\includegraphics[width=0.48\textwidth]{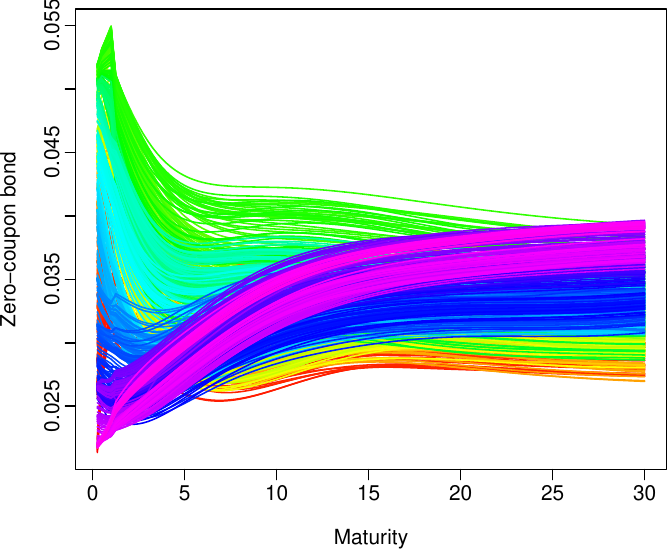}}
\quad
\subfloat[Functional Autocorrelation plot]
{\includegraphics[width=0.48\textwidth]{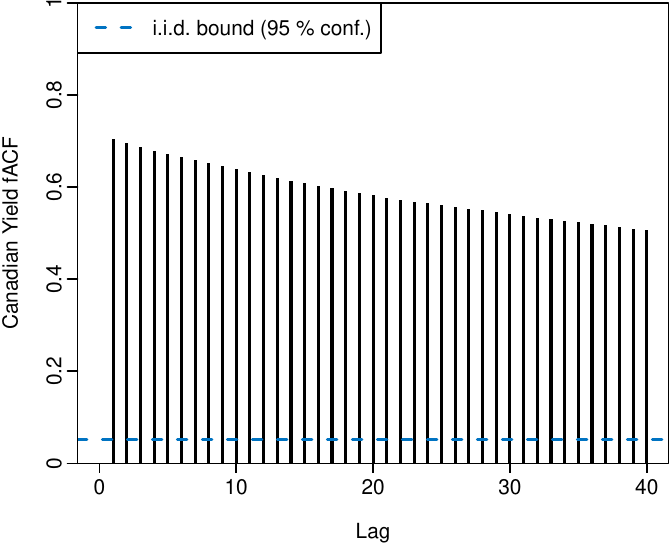}}
\caption{\small Graphical displays of end-of-day Canadian yield curves from 4 July 2022 to 31 December 2025.}
\label{fig:8}
\end{figure}

In Figure~\ref{fig:8}, we present a rainbow plot and a functional autocorrelation plot of the Canadian yield curve data across maturities. The curves display substantial persistence over time, with neighbouring daily yield curves exhibiting highly similar shapes and gradual temporal evolution. Such behaviour is consistent with the widely documented persistence and nonstationary characteristics of yield curve dynamics.

\begin{figure}[!htb]
\centering
\includegraphics[width=0.85\textwidth]{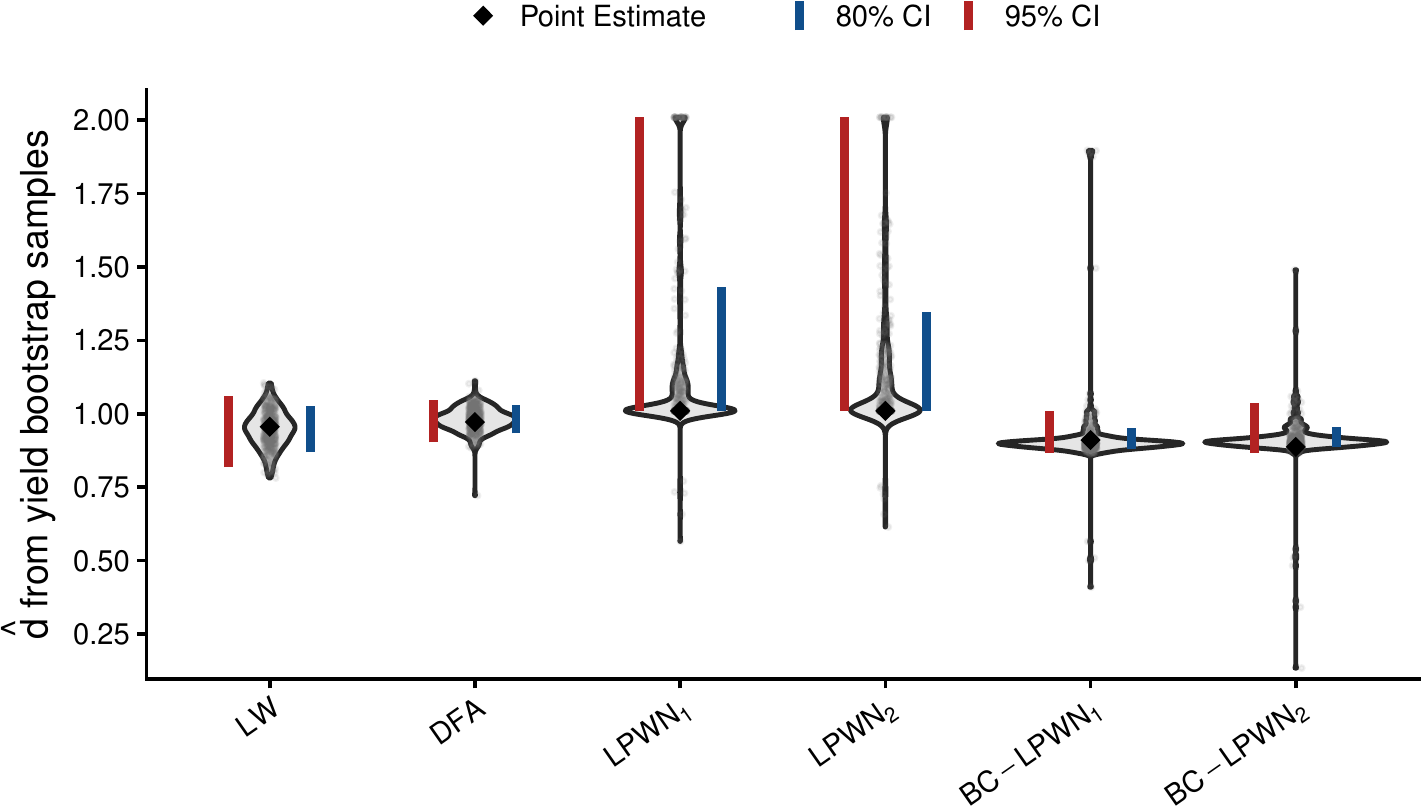}
\caption{\small Bootstrap distributions for estimated $d$ with the associated empirical 80\% and 95\% confidence intervals across each estimator with Canadian yield data.}
\label{fig:9}
\end{figure}

Figure~\ref{fig:9} further displays the bootstrap distributions together with the nominal 80\% and 95\% confidence intervals for the estimators. Overall, all estimators produce memory parameter estimates close to one, matching the empirical evidence that the Canadian yield curve process behaves approximately as an \(I(1)\) functional time series. Similar to the mortality analysis, LPWN$_{1}$ and LPWN$_{2}$ exhibit visible accumulation of bootstrap estimates near values such as $1.01$ and $2.01$, arising from the iterative differencing reconstruction procedure under the admissible parameter range constraint. As the bootstrap bias-corrected estimators BC-LPWN$_1$ and BC-LPWN$_2$ substantially alleviate this boundary concentration effect and yield smoother bootstrap distributions, it potentially decreases variability due to more stable, less boundary-constrained bootstrap estimates.

\section{Conclusion}\label{sec:7}

We investigated bootstrap bias correction for estimating the memory parameter $d$ in fractionally integrated functional time series. Motivated by the substantial finite-sample bias observed in classical semiparametric estimators (e.g., LW, DFA) under strong short-range dependence, we proposed a prefiltered sieve bootstrap procedure built upon the LPWN estimator. The proposed approach combines the flexibility of the polynomial local Whittle framework with the resampling feature of the sieve bootstrap, allowing for further reduction of estimation bias in the presence of strong autoregressive dependence.

In FARFIMA$(1,d,0)$, FARFIMA$(0,d,1)$, and FARFIMA$(1,d,1)$ settings, via simulation studies, we demonstrated that the proposed BC-LPWN provides substantial bias reduction relative to traditional semiparametric estimators, particularly under strong short-range dependence. While the LPWN estimators already mitigate the bias induced by strong autoregressive dependence, the use of sieve-bootstrap further improves the LPWN estimator across a broad range of settings. In addition, the proposed bootstrap procedure naturally provides confidence intervals for memory parameter estimation through the bootstrap samples, offering a practical uncertainty quantification framework alongside point estimation.

We further illustrated the practical relevance of the proposed methodology through empirical applications to Swedish male and female mortality rates and Canadian yield curve data. The empirical results demonstrate that the proposed bootstrap framework provides meaningful point and interval estimation, aligning with existing literature \citep[c.f.,][]{LRS23, SS26}, for strongly persistent functional time series frequently encountered in demographic and financial studies.

There are several ways in which the current methodology can be further extended, and we briefly mention a few:
\begin{asparaenum}
\item[1)] The simulation results indicate that the relative performance of competing estimators may depend on the strength of the underlying short-range dependence. This suggests the potential development of unified procedures that jointly estimate the degree of short-range dependence and select the appropriate memory estimator accordingly.
\item[2)] While the present work primarily focuses on point estimation for the memory parameter in simulation studies, a more comprehensive investigation of estimation variability, including empirical confidence intervals and coverage probability under varying dependence structures, is warranted. It may provide further insight into the finite-sample behaviour of the proposed estimators.
\end{asparaenum}

\section*{Acknowledgment}

The first author gratefully acknowledges the financial support provided through the International Macquarie University Research Excellence Scholarship. The second author is grateful for financial support from an Australian Research Council Future Fellowship (FT240100338). We are grateful for the comments provided by Prof. Liudas Giraitis from the Queen Mary University of London and Dr. Won-Ki Seo from the University of Sydney.

\section*{Supplement}

\begin{description}
\item[Code for bias-corrected estimator with sieve bootstrapping.] The \Rlogo \ code to estimate the long-memory parameter from the proposed approaches described in the manuscript. The \Rlogo \ code is available at a \href{https://github.com/chang-liu-0122/Bias-Corrected-Memory}{GitHub repository}.
\item[Interface and code for shiny application.] A shiny user interface for visualising and analysing simulation results in different configurations is available on the dashboard of the \href{ https://changliu0122.shinyapps.io/BC-SIM/}{Shiny application}. The \Rlogo \ code for the Shiny application is also available at the \href{https://github.com/chang-liu-0122/Bias-Corrected-Memory/blob/main/Simulation/bootstrap%20bias%20correction%20shiny.R}{GitHub repository}.
\end{description}

\newpage
\bibliographystyle{agsm}
\bibliography{BC_memory_LPWN_sieve}

\clearpage

%supplement

\begin{appendices}

\section{Supplementary results for additional figures}\label{sec:S1}

\begin{figure}[H]
\centering
\includegraphics[width=0.9\textwidth]{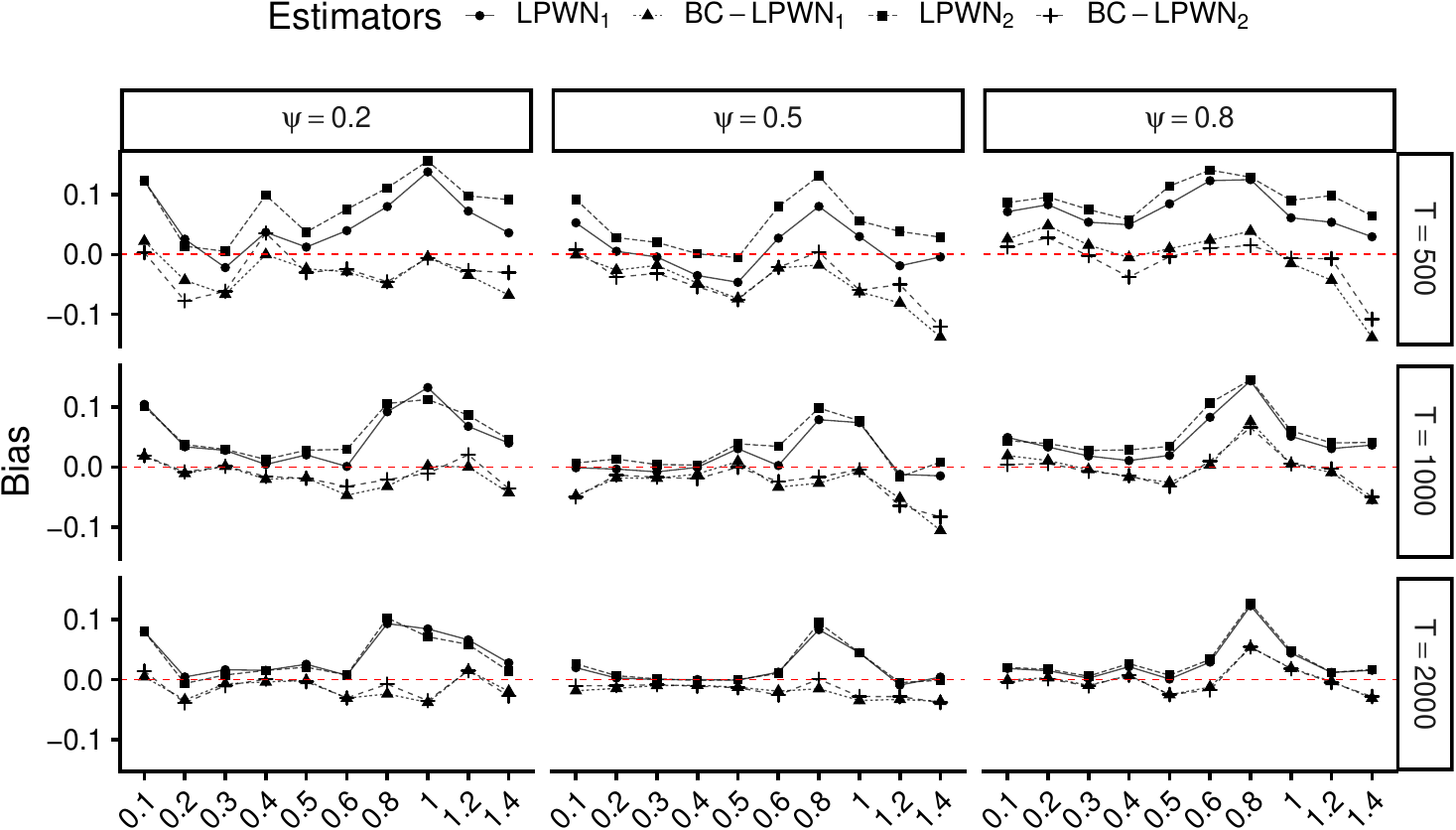}\\
\vspace{.3in}
\includegraphics[width=0.9\textwidth]{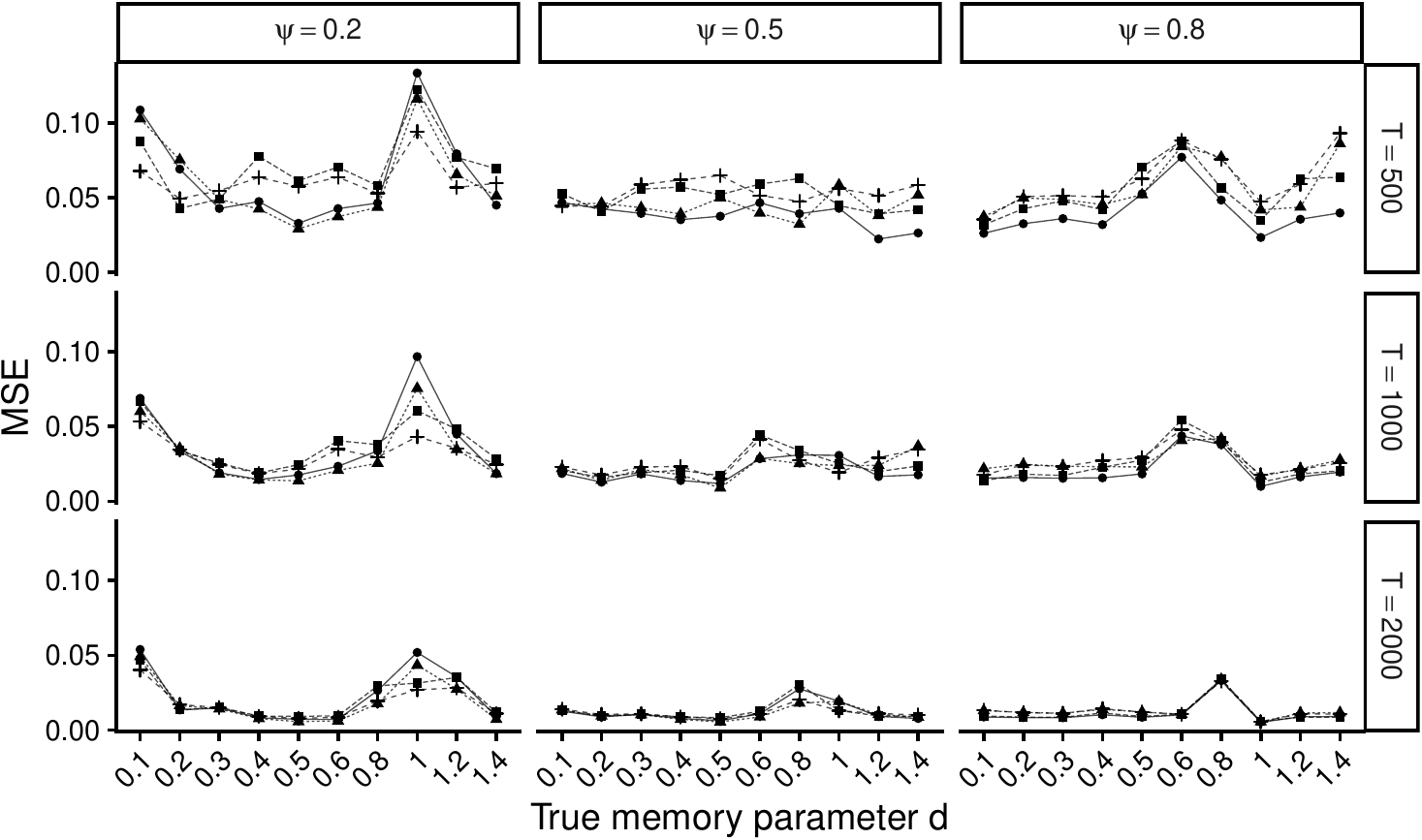}
\caption{\small Performance comparison of LPWN$_1$, LPWN$_2$, BC-LPWN$_1$, and BC-LPWN$_2$ the $FARFIMA(1,d,0)$ setting across different values of memory parameter $d$, sample size $T$, and autoregressive strength $\psi$.}
\label{fig:sup1}
\end{figure}

\begin{figure}[H]
\centering
\includegraphics[width=0.9\textwidth]{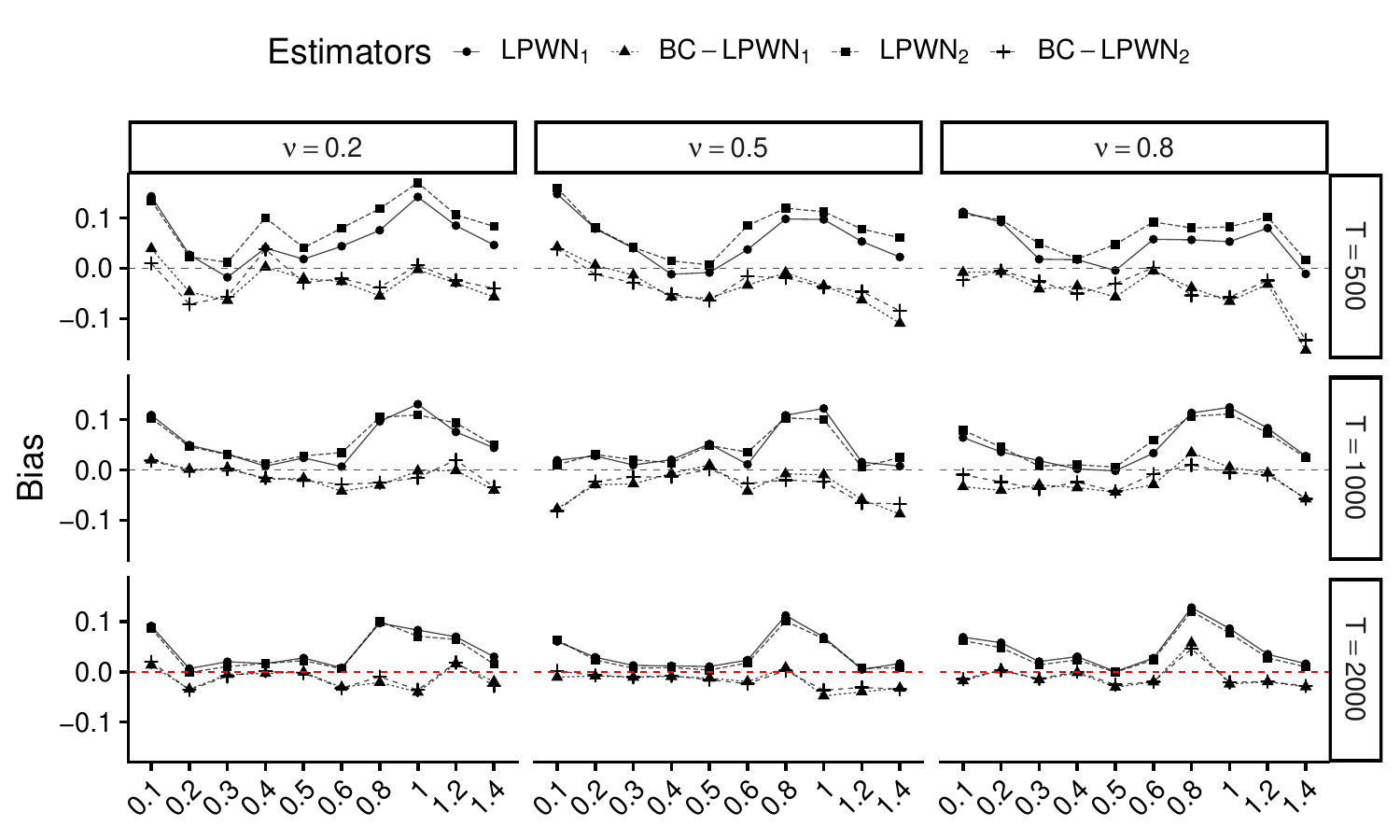}\\
\vspace{.3in}
\includegraphics[width=0.9\textwidth]{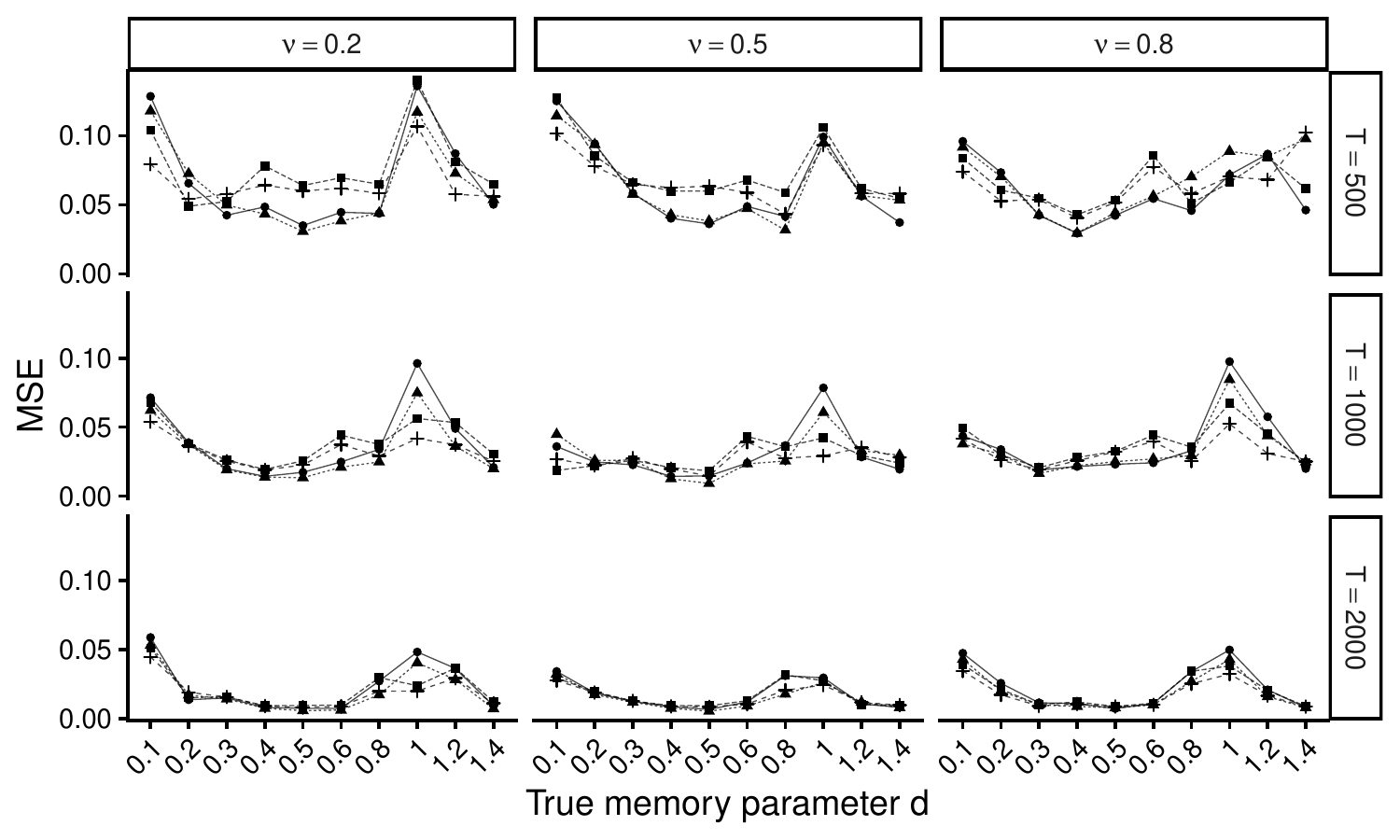}
\caption{\small Performance comparison of LPWN$_1$, LPWN$_2$, BC-LPWN$_1$, and BC-LPWN$_2$ the $FARFIMA(0,d,1)$ setting across different values of $d$, $T$, and $\nu$.}
\label{fig:sup2}
\end{figure}

\begin{figure}[H]
\centering
\includegraphics[width=0.9\textwidth]{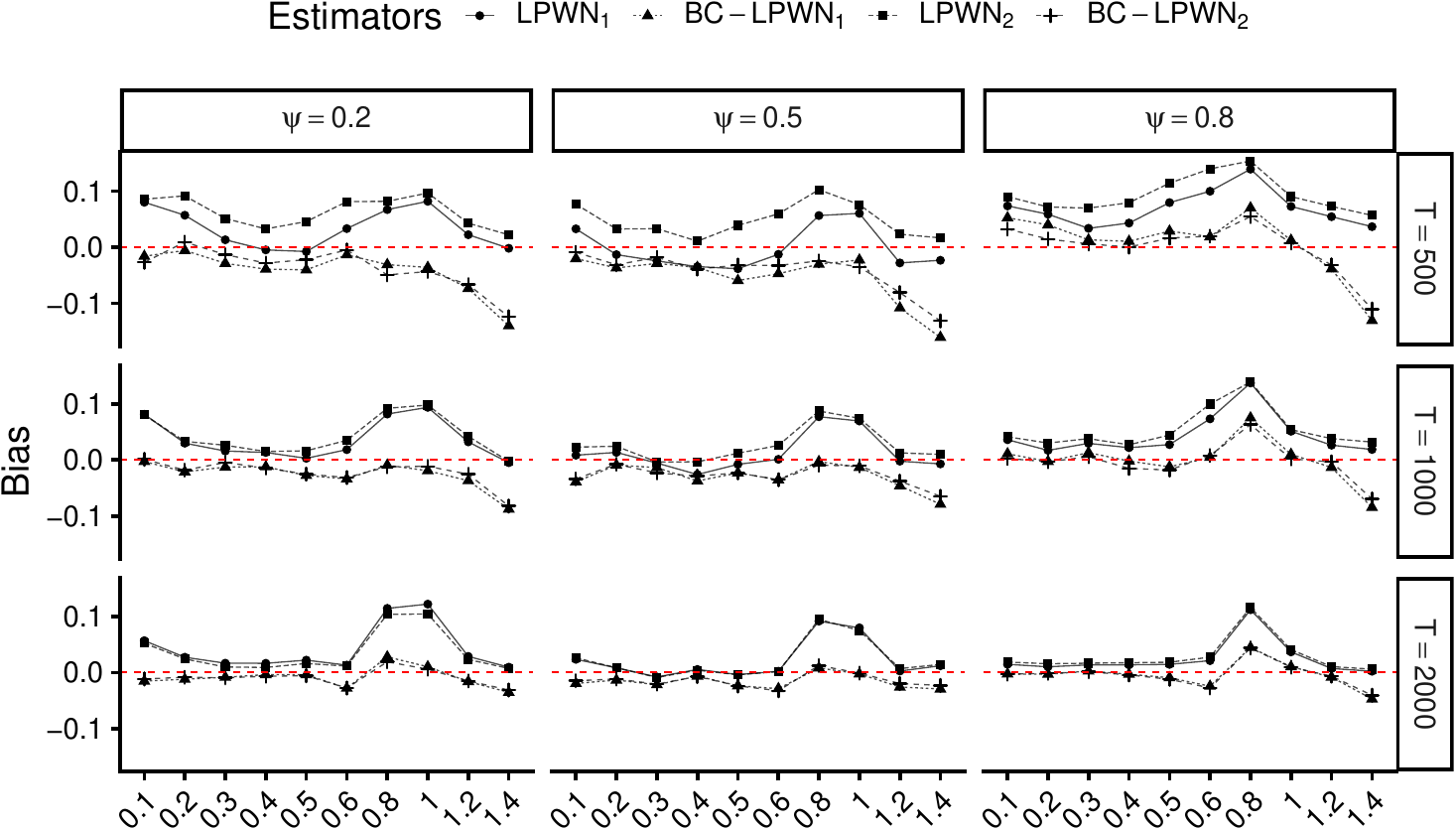} \\
\vspace{.3in}
\includegraphics[width=0.9\textwidth]{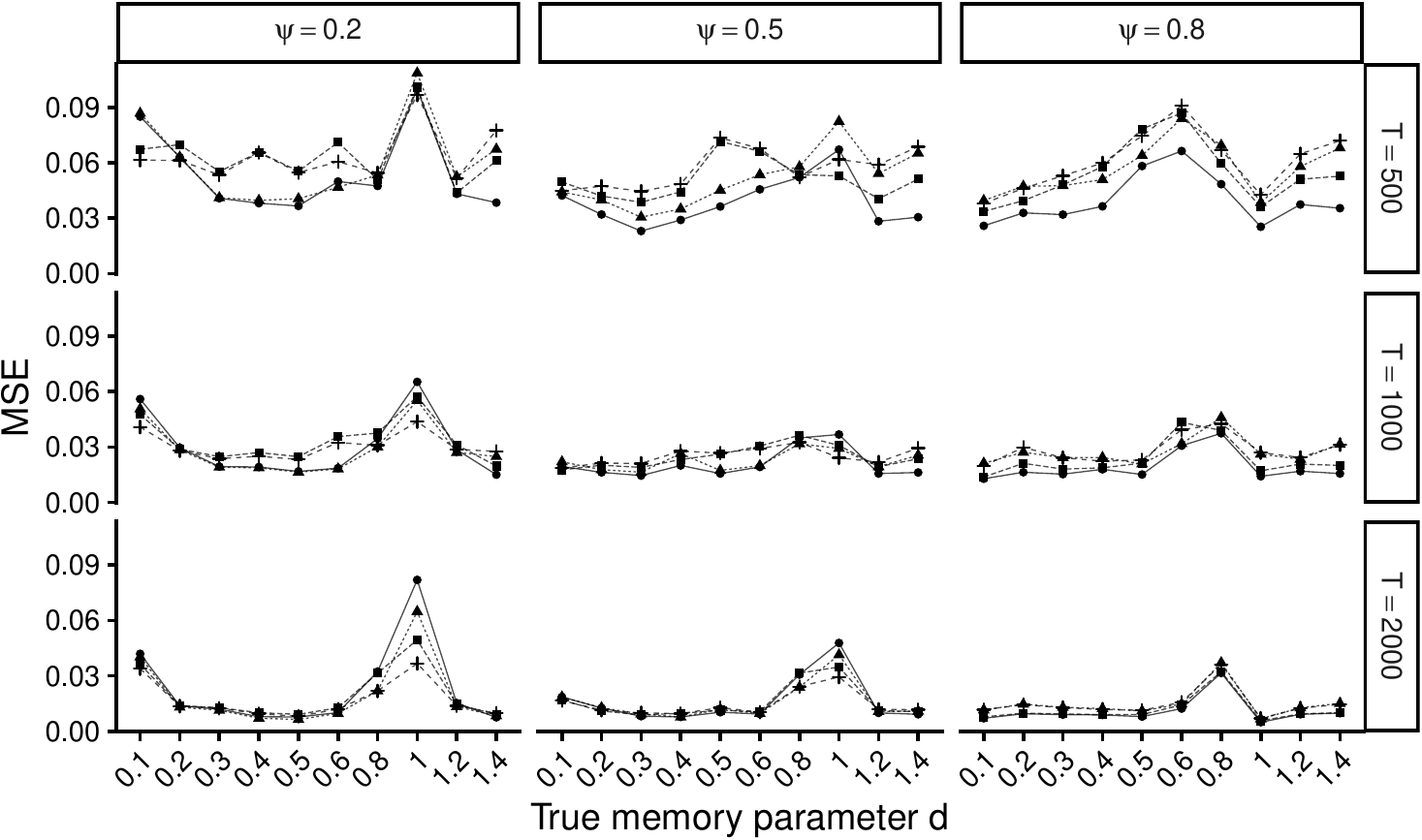}
\caption{\small Performance comparison of LPWN$_1$, LPWN$_2$, BC-LPWN$_1$, and BC-LPWN$_2$ the $FARFIMA(1,d,1)$ setting across different values of $d$, $T$, and $\psi$.}
\label{fig:sup3}
\end{figure}

Figures~\ref{fig:sup1}, \ref{fig:sup2} and \ref{fig:sup3} provide additional comparisons among different variants of the LPWN-based and BC-LPWN-based estimators. Specifically, the figures compare the first- and second-order LPWN estimators, LPWN$_1$ and LPWN$_2$, together with their sieve-bootstrap bias-corrected versions, BC-LPWN$_1$ and BC-LPWN$_2$. These comparisons are conducted across different values of the memory parameter $d$, sample size $T$, and autoregressive strength $\psi$. 
\begin{itemize}
\item Figure~\ref{fig:sup1} reports the comparison under the FARFIMA$(1,d,0)$ setting;
\item Figure~\ref{fig:sup2} reports the comparison under the FARFIMA$(0,d,1)$ setting; 
\item Figure~\ref{fig:sup3} reports the corresponding comparison under the FARFIMA$(1,d,1)$ setting. 
\end{itemize}
These supplementary figures are included to examine how the polynomial order in the LPWN estimator and the additional bootstrap bias correction affect estimation performance under different DGPs.

\newpage
\section{Supplementary results for additional tables}\label{sec:S2}

Tables~\ref{tab:S1}--\ref{tab:S6} provide the numerical results corresponding to the supplementary estimator comparisons. Specifically, the tables report the average bias and MSE of LW, DFA, LPWN$_1$, LPWN$_2$, BC-LPWN$_1$, and BC-LPWN$_2$ over 100 simulation repetitions. The comparisons are conducted across different values of the memory parameter $d$, sample size $T$, and short-run dependence strengths $(\psi, \nu)$. Specifically, Tables~\ref{tab:S1} and~\ref{tab:S2} report the bias and MSE results under the FARFIMA$(1,d,0)$ setting, respectively, where the short-run dependence is controlled by $\psi$. 
\begin{small}
\tabcolsep 0.2in
\renewcommand{\arraystretch}{0.85}
\begin{longtable}{@{}lrrrrrrrr@{}} 
\caption{Averaged over 100 repetitions, we present the simulation results in terms of bias for $T=500, 1000, 2000$ and $\psi = 0.2, 0.5, 0.8$ among different estimators. The estimator with the smallest absolute bias is highlighted in bold.}
\label{tab:S1} \\
\toprule
    &   & & \multicolumn{6}{c}{Bias} \\
 $T$ & $\psi$ & $d$ & LW & DFA & LPWN$_1$ & LPWN$_2$ & BC-LPWN$_1$ & BC-LPWN$_2$\\
\midrule
\endfirsthead
\caption[]{Simulation results for bias comparisons (continued).} \\
\toprule
 $T$ & $\psi$ & $d$ & LW & DFA & LPWN$_1$ & LPWN$_2$ & BC-LPWN$_1$ & BC-LPWN$_2$ \\
\midrule
\endhead
\midrule
\multicolumn{9}{r}{Continued on next page}
\endfoot
\endlastfoot
500 & 0.2 & 0.1 & -0.0284 & 0.0080 & 0.1231 & 0.1235 & 0.0222 & \textBF{0.0033} \\ 
&  & 0.2 & -0.0265 & \textBF{-0.0048} & 0.0256 & 0.0139 & -0.0431 & -0.0773\\ 
&  & 0.3 & -0.0188 & -0.0147 & -0.0218 & \textbf{0.0053} & -0.0664 & -0.0614 \\ 
&  & 0.4 & -0.0009 & -0.0225 & 0.0371 & 0.0991 & \textbf{-0.0002} & 0.0358\\ 
&  & 0.5 & \textbf{-0.0103} & -0.0284 & 0.0124 & 0.0375 & -0.0239 & -0.0301\\ 
\cmidrule{3-9}
&  & 0.6 & \textbf{-0.0024} & -0.0326 & 0.0399 & 0.0754 & -0.0287 & -0.0240 \\ 
&  & 0.8 & -0.0140 & \textbf{0.0088} & 0.0801 & 0.1112 & -0.0498 & -0.0455 \\ 
&  & 1.0 & -0.0175 & 0.0256 & 0.1380 & 0.1562 & \textbf{-0.0043} & -0.0056 \\
\bottomrule
\end{longtable}
\end{small}

\begin{small}
\tabcolsep 0.2in
\renewcommand{\arraystretch}{0.93}
\begin{longtable}{@{}lrrrrrrrr@{}}
\caption{Averaged over 100 repetitions, we present the simulation results in terms of MSE for $T=500, 1000, 2000$ and $\psi = 0.2, 0.5, 0.8$ among different estimators. The estimator with the smallest MSE is highlighted in bold.}
\label{tab:S2} \\
\toprule
    & &   & \multicolumn{6}{c}{MSE} \\
 $T$ & $\psi$ & $d$ & LW & DFA & LPWN$_1$ & LPWN$_2$ & BC-LPWN$_1$ & BC-LPWN$_2$ \\
\midrule
\endfirsthead
\caption[]{Simulation results for MSE comparisons (continued).} \\
\toprule
 $T$ & $\psi$ & $d$ & LW & DFA & LPWN$_1$ & LPWN$_2$ & BC-LPWN$_1$ & BC-LPWN$_2$ \\
\midrule
\endhead
\midrule
\multicolumn{9}{r}{Continued on next page}
\endfoot
\endlastfoot
500 & 0.2 & 0.1 & 0.0078 & \textbf{0.0027} & 0.1086 & 0.0876 & 0.1027 & 0.0677 \\ 
   &  & 0.2 & 0.0055 & \textbf{0.0031} & 0.0691 & 0.0428 & 0.0753 & 0.0493 \\ 
  &  & 0.3 & 0.0071 & \textbf{0.0037} & 0.0428 & 0.0491 & 0.0490 & 0.0547 \\ 
  &  & 0.4 & 0.0059 & \textbf{0.0044} & 0.0474 & 0.0776 & 0.0424 & 0.0635 \\ 
  &  & 0.5 & 0.0056 & \textbf{0.0051} & 0.0329 & 0.0614 & 0.0292 & 0.0576 \\ 
  \cmidrule{3-9}
  &  & 0.6 & \textbf{0.0052} & 0.0058 & 0.0427 & 0.0704 & 0.0373 & 0.0637 \\ 
  &  & 0.8 & \textbf{0.0057} & 0.0104 & 0.0464 & 0.0580 & 0.0436 & 0.0529 \\ 
 &  & 1.0 & 0.0058 & \textbf{0.0030} & 0.1333 & 0.1223 & 0.1160 & 0.0942  \\
\bottomrule
\end{longtable}
\end{small}

\clearpage

Tables~\ref{tab:S3} and~\ref{tab:S4} report the bias and MSE results under the FARFIMA$(0,d,1)$ setting, respectively, where the short-run dependence is controlled by $\nu$.
\begin{small}
\tabcolsep 0.2in
\renewcommand{\arraystretch}{0.912}
\begin{longtable}{@{}lrrrrrrrr@{}}
\caption{Averaged over 100 repetitions, we present the simulation results in terms of bias for $T=500, 1000, 2000$ and $\nu=0.2, 0.5, 0.8$ among different estimators. The estimator with the smallest absolute bias is highlighted in bold.}
\label{tab:S3} \\
\toprule
 & & & \multicolumn{6}{c}{Bias} \\
$T$ & $\nu$ & $d$ & LW & DFA & LPWN$_1$ & LPWN$_2$ & BC-LPWN$_1$ & BC-LPWN$_2$ \\
\midrule
\endfirsthead
\caption[]{Simulation results for bias comparisons under the FARFIMA$(0,d,1)$ setting (continued).} \\
\toprule
$T$ & $\nu$ & $d$ & LW & DFA & LPWN$_1$ & LPWN$_2$ & BC-LPWN$_1$ & BC-LPWN$_2$ \\
\midrule
\endhead
\midrule
\multicolumn{9}{r}{Continued on next page} \\
\endfoot
\endlastfoot
500 & 0.2 & 0.1 & -0.0373 & \textbf{0.0002} & 0.1430 & 0.1345 & 0.0391 & 0.0097 \\
 &  & 0.2 & -0.0354 & \textbf{-0.0046} & 0.0268 & 0.0227 & -0.0470 & -0.0714 \\
 &  & 0.3 & -0.0278 & -0.0228 & -0.0179 & \textbf{0.0119} & -0.0639 & -0.0562 \\
 &  & 0.4 & -0.0092 & -0.0247 & 0.0401 & 0.1004 & \textbf{0.0025} & 0.0375 \\
 &  & 0.5 & -0.0193 & -0.0264 & \textbf{0.0182} & 0.0407 & -0.0204 & -0.0287 \\
\cmidrule{3-9}
 &  & 0.6 & \textbf{-0.0098} & -0.0309 & 0.0438 & 0.0799 & -0.0264 & -0.0208 \\
 &  & 0.8 & -0.0229 & \textbf{-0.0074} & 0.0755 & 0.1183 & -0.0550 & -0.0378 \\
 &  & 1.0 & -0.0262 & 0.0157 & 0.1417 & 0.1701 & \textbf{-0.0029} & 0.0062 \\
\bottomrule
\end{longtable}
\end{small}

\begin{small}
\tabcolsep 0.2in
\renewcommand{\arraystretch}{0.93}
\begin{longtable}{@{}lrrrrrrrr@{}}
\caption{Averaged over 100 repetitions, we present the simulation results in terms of MSE for $T=500, 1000, 2000$ and $\nu=0.2, 0.5, 0.8$ among different estimators. The estimator with the smallest MSE is highlighted in bold.}
\label{tab:S4} \\
\toprule
 & & & \multicolumn{6}{c}{MSE} \\
$T$ & $\nu$ & $d$ & LW & DFA & LPWN$_1$ & LPWN$_2$ & BC-LPWN$_1$ & BC-LPWN$_2$ \\
\midrule
\endfirsthead
\caption[]{Simulation results for mse comparisons under the FARFIMA$(0,d,1)$ setting (continued).} \\
\toprule
$T$ & $\nu$ & $d$ & LW & DFA & LPWN$_1$ & LPWN$_2$ & BC-LPWN$_1$ & BC-LPWN$_2$ \\
\midrule
\endhead
\midrule
\multicolumn{9}{r}{Continued on next page} \\
\endfoot
\endlastfoot
500 & 0.2 & 0.1 & 0.0083 & \textbf{0.0038} & 0.1286 & 0.1041 & 0.1180 & 0.0795 \\
 &  & 0.2 & 0.0060 & \textbf{0.0034} & 0.0656 & 0.0490 & 0.0726 & 0.0542 \\
 &  & 0.3 & 0.0075 & \textbf{0.0045} & 0.0425 & 0.0524 & 0.0498 & 0.0579 \\
 &  & 0.4 & 0.0061 & \textbf{0.0053} & 0.0485 & 0.0782 & 0.0433 & 0.0643 \\
 &  & 0.5 & 0.0058 & \textbf{0.0045} & 0.0350 & 0.0640 & 0.0307 & 0.0601 \\
\cmidrule{3-9}
 &  & 0.6 & \textbf{0.0054} & 0.0062 & 0.0446 & 0.0696 & 0.0383 & 0.0619 \\
 &  & 0.8 & \textbf{0.0061} & 0.0095 & 0.0438 & 0.0649 & 0.0441 & 0.0584 \\
 &  & 1.0 & 0.0062 & \textbf{0.0042} & 0.1361 & 0.1402 & 0.1172 & 0.1069 \\
\bottomrule
\end{longtable}
\end{small}

\clearpage

Tables~\ref{tab:S5} and~\ref{tab:S6} report the corresponding bias and MSE results under the FARFIMA$(1,d,1)$ setting, where the short-run dependence is controlled by $(\psi,\nu)$.

\begin{small}
\tabcolsep 0.17in
\renewcommand{\arraystretch}{0.913}
\begin{longtable}{@{}lllrrrrrrr@{}}
\caption{Averaged over 100 repetitions, we present the simulation results in terms of bias for $T=500,1000,2000$ under $(\psi,\nu)=(0.2,0.2),(0.5,0.5),(0.8,0.8)$ among different estimators. The estimator with the smallest absolute bias is highlighted in bold.}\label{tab:S5} \\
\toprule
& & & & \multicolumn{6}{c}{Bias} \\
$T$ & ($\psi$, $\nu$) & $d$ & LW & DFA & LPWN$_1$ & LPWN$_2$ & BC-LPWN$_1$ & BC-LPWN$_2$ \\
\midrule
\endfirsthead
\caption[]{Simulation results for bias comparisons (continued).} \\
\toprule
$T$ & ($\psi$, $\nu$) & $d$
& LW & DFA & LPWN$_1$ & LPWN$_2$ & BC-LPWN$_1$ & BC-LPWN$_2$ \\
\midrule
\endhead
\midrule
\multicolumn{10}{r}{Continued on next page}
\endfoot
\endlastfoot
500  & (0.2, 0.2) & 0.1 & -0.0202 & 0.0385 & 0.1117 & 0.1185 & 0.0226 & \textbf{0.0080}  \\ 
&     & 0.2 & \textbf{0.0037} & 0.0230 & 0.0915 & 0.1370 & 0.0407 & 0.0597   \\ 
&     & 0.3 & \textbf{-0.0073} & 0.0102 & 0.0105 & 0.0487 & -0.0198 & -0.0074  \\ 
&     & 0.4 & 0.0125 & \textbf{-0.0004} & 0.0192 & 0.0584 & -0.0038 & 0.0020  \\ 
&     & 0.5 & 0.0087 & -0.0091 & \textbf{0.0080} & 0.0544 & -0.0303 & -0.0204  \\ 
\cmidrule{3-10}
&     & 0.6 & 0.0189 & -0.0160 & 0.0307 & 0.0982 & -0.0182 & \textbf{0.0107}  \\
&     & 0.8 & \textbf{-0.0127} & 0.0441 & 0.0688 & 0.0880 & -0.0432 & -0.0540  \\ 
&     & 1.0 & \textbf{0.0072} & 0.0577 & 0.0927 & 0.1011 & -0.0310 & -0.0408    \\
\bottomrule
\end{longtable}
\end{small}

\begin{small}
\tabcolsep 0.17in
\renewcommand{\arraystretch}{0.913}
\begin{longtable}{@{}lllrrrrrrr@{}}
\caption{Averaged over 100 repetitions, we present the simulation results in terms of mse for $T=500,1000,2000$ under $(\psi,\nu)=(0.2,0.2),(0.5,0.5),(0.8,0.8)$ among different estimators. The estimator with the smallest MSE is highlighted in bold.}
\label{tab:S6} \\
\toprule
& & & \multicolumn{6}{c}{MSE} \\
$T$ & ($\psi$, $\nu$) & $d$
& LW & DFA & LPWN$_1$ & LPWN$_2$ & BC-LPWN$_1$ & BC-LPWN$_2$  \\
\midrule
\endfirsthead
\caption[]{Simulation results for MSE comparisons (continued).} \\
\toprule
$T$ & ($\psi$, $\nu$) & $d$
& LW & DFA & LPWN$_1$ & LPWN$_2$ & BC-LPWN$_1$ & BC-LPWN$_2$ \\
\midrule
\endhead
\midrule
\multicolumn{9}{r}{Continued on next page}
\endfoot
\endlastfoot
  500  & (0.2, 0.2) & 0.1   & 0.0074 &\textbf{ 0.0042} & 0.1024 & 0.0861 & 0.0980 & 0.0705\\
       &     & 0.2   & 0.0057 & \textbf{0.0036} & 0.0801 & 0.0915 & 0.0753 & 0.0735  \\
       &     & 0.3   & 0.0052 & \textbf{0.0036} & 0.0388 & 0.0538 & 0.0381 & 0.0541 \\
       &     & 0.4   & 0.0058 & \textbf{0.0039} & 0.0471 & 0.0747 & 0.0443 & 0.0668  \\
       &     & 0.5   & 0.0076 & \textbf{0.0045} & 0.0435 & 0.0574 & 0.0479 & 0.0565  \\
\cmidrule{3-10}
       &     & 0.6   & 0.0068 & \textbf{0.0051} & 0.0400 & 0.0728 & 0.0306 & 0.0567 \\
       &     & 0.8   & \textbf{0.0061} & 0.0147 & 0.0426 & 0.0457 & 0.0411 & 0.0466 \\
       &     & 1.0   & \textbf{0.0051} & 0.0057 & 0.0997 & 0.0933 & 0.0979 & 0.0816 \\
\bottomrule
\end{longtable}
\end{small}

These supplementary tables are included to provide detailed numerical values for the graphical comparisons and to support the assessment of estimator performance across different DGPs.

\end{appendices}

\end{document}